\documentclass[fleqn,usenatbib]{mnras}
\usepackage{newtxtext,newtxmath}
\usepackage{txfonts}
\usepackage[T1]{fontenc}
\usepackage{soul}
\usepackage{float}

\DeclareRobustCommand{\VAN}[3]{#2}
\let\VANthebibliography\thebibliography
\def\thebibliography{\DeclareRobustCommand{\VAN}[3]{##3}\VANthebibliography}

\usepackage{graphicx}	% Including figure files
\usepackage{multirow}
\usepackage{comment}

\usepackage[mathscr]{euscript}
\usepackage[T1]{fontenc}
\usepackage{hyperref}

\title{Deciphering the properties of UV upturn galaxies in the Virgo cluster}
\author[K. R. Akhil]{
Krishna R. Akhil \thanks{E-mail:akhil.r@res.christuniversity.in}$^{1}$, Sreeja S Kartha \thanks{E-mail:sreeja.kartha@christuniversity.in}$^{1}$, Namitha Kizhuprakkat$^{2}$, K. Ujjwal$^{1}$, Niranjana P$^{1}$
\\
$^{1}$Department of Physics and Electronics, CHRIST (Deemed to be University), Bangalore 560029, India
\\
$^{2}$Institute of Astronomy and Department of Physics, National Tsing Hua University, 101 Kuang-Fu Rd. Sec. 2, Hsinchu 30013, Taiwan\\}

\date{Accepted 2024 October 14. Received 2024 August 23; in original form 2023 September 30}

\pubyear{2023}
\begin{document}
\label{firstpage}
\pagerange{\pageref{firstpage}--\pageref{lastpage}}
\maketitle
\begin{abstract}
The UV upturn refers to the increase in UV flux at wavelengths shorter than 3000 $\AA$ observed in quiescent early-type galaxies (ETGs), which still remains a puzzle. In this study, we aim to identify ETGs showing the UV upturn phenomenon within the Virgo galaxy cluster. We utilized a color-color diagram to identify all potential possible UV upturn galaxies. The Spectral Energy Distributions (SED) of these galaxies were then analyzed using the CIGALE software; we confirmed the presence of UV upturn in galaxies within the Virgo cluster. We found that the SED fitting method is the best tool to visualize and confirm the UV upturn phenomenon in ETGs. Our findings reveal that the population distributions regarding stellar mass and star formation rate properties are similar between UV upturn and red sequence galaxies. We suggest that the UV contribution originates from old stellar populations and can be modeled effectively without a burst model. Moreover, by estimating the temperature of the stellar population responsible for the UV emission, we determined it to be 13,000 K to 18,000 K. These temperature estimates support the notion that the UV upturn likely arises from the contribution of low mass evolved stellar populations (extreme horizontal branch stars). Furthermore, the Mg2 index, a metallicity indicator, in the confirmed upturn galaxies shows higher strength and follows a similar trend to previous studies. This study sheds light on the nature of UV upturn galaxies within the Virgo cluster and provides evidence that low-mass evolved stellar populations are the possible mechanisms driving the UV upturn phenomenon.
\end{abstract}

%%%%%%%%%%%%%%%%% BODY OF PAPER %%%%%%%%%%%%%%%%%%
\section{Introduction}
The excess in flux exhibited by a significant number of Early Type Galaxies (ETGs) at $\lambda$ $<$  3000 $\AA$, compared to the expected contributions from their old, metal-rich stellar populations, is known as Ultraviolet upturn (hereafter UV upturn) \citep{Code1979, Bertola1980, Oconnel1986}. A consensus regarding the formation of the UV upturn phenomenon is yet to be achieved. Initially, the residual star formation activity in ETGs \citep[e.g.][]{Evans2018} was considered to be the cause of the UV upturn phenomenon \citep[e.g.][]{Yi2005, Kaviraj2007,sheen2016}. Recent studies have introduced alternative contributors to the UV upturn, including binary star systems and planetary nebulae as potential contributors to the UV upturn \citep{2013Pastorello,2014perezandbruzual}. However, the widely accepted theory considers the contribution from helium-rich evolved stellar populations like extreme horizontal branch (EHB) stars and post--asymptotic giant branch (P-AGB) stars to be the origin of the UV upturn phenomenon \citep{Greggioandrezini1990, Ferguson1992, Yi1997, Brown1998, OConnell1999, Brown2000, Yoon2004, Han2007, peng&nagai2009, Schombert2016}.

While \citet{Burstein1988} initially reported the presence of the upturn property in nearby isolated elliptical galaxies, subsequent research has predominantly concentrated on studying ETGs that are considered prominent candidate members within nearby clusters like Virgo \citep{Boselli2005} and Coma \citep{Smith2012}. Recognizing the intricacy of this phenomenon and the necessity to comprehensively identify all potential upturn candidates within a cluster of galaxies, we propose the Virgo cluster as the ideal template for our study. The Virgo cluster, due to its proximity and the wealth of multiwavelength data from dedicated surveys such as the GALEX UltraViolet Virgo Cluster Survey \citep[GUViCS;][]{boselli2011}, A Virgo Environmental Survey Tracing Ionised Gas Emission \citep[VESTIGE;][]{2018BoselliVestige}, The Herschel Virgo Cluster Survey \citep[HeViCS;][]{2010Davies_Hevics}, stands out as one of the best templates for attempting to understand the properties of upturn galaxies.\par

The major challenge in the studies related to the galaxies with UV upturn lies in the identification of candidate galaxies. The best method to determine the quiescent galaxies exhibiting UV upturn is to explore the spectra in the UV region. The limited availability of UV spectra constrains the identification of only a limited number of galaxies. 
On the other hand, the wealth of multi-wavelength photometric data opened a new window of opportunity to identify the possible upturn galaxies.  \citet{Burstein1988}, \citet{ Boselli2005}, \citet{Yi2011}, \citet{Donas2007}, and \citet{philips2020} introduced and studied different criteria for identifying UV upturn galaxies based on  UV and optical colors.  These studies explored the hypothesis that the UV upturn is driven by hot, evolved stars. In addition, \citet{Ali2018a} utilized GALEX and UVOT data, specifically below 3000 Å, to construct UV SEDs for galaxies in the Coma cluster. Similarly, \citet{Burstein1988} focused on the ultraviolet SEDs of UV upturn galaxies within the Virgo cluster. SED preserves the imprints of the baryonic processes that result in the formation and evolution of the galaxy. The SED-fitting methods are the best tool to extract information from a galaxy. Numerous modules have been created over the years to model the panchromatic emission from galaxies. Relying on the energy balancing concept, i.e., the energy emitted by dust in the IR precisely corresponds to the energy absorbed by dust in the UV-optical range, is becoming the best compromise in terms of speed, precision, and accuracy. The energy balancing concept is the base of the widely used SED modeling codes such as CIGALE \citep{Burgarella2005, Noll2009, Boquien2019},
MAGPHYS \citep{daCunha2008}, and FSPS \citep{Conroy2009, Conroy2010}. The parameters obtained from the SED-fitting of a well-characterized sample of UV upturn galaxies will provide more insights regarding the properties of the upturn galaxies. \par

In this context, several studies have identified and examined the properties of UV upturn galaxies using various selection criteria and observed spectra. However, it remains unclear whether the UV upturn in these galaxies is solely due to old stellar populations or if contributions from young stellar populations are also involved. To better understand this, we intend to use CIGALE, which incorporates different models, including those for star formation, stellar populations, and dust, to separately analyze the stellar contributions in UV upturn galaxies. Further, we also aim to understand the role of evolved stars in the UV upturn phenomenon. The paper is arranged as follows. Data collection is explained in section \ref{sec:data}. Methodologies to distinguish quiescent ETGs from Star-forming Galaxies (SFGs), identification of upturn galaxies followed by confirmation of UV upturn are included in section \ref{sec:identific}. We discuss our observations in section \ref{sec:discuss} and a summary in section \ref{sec:summaty}.

\section{Data Inventory}
\label{sec:data}
Our study focuses on ETGs of the Virgo cluster, which is one of the closest galaxy clusters to our local group at a mean redshift of 0.0036 \citep{Ebeling1998}. We have compiled the previously identified Virgo cluster galaxies from Virgo Cluster Catalog \cite[VCC,][]{Binggeli1985} and its modified version; Extended Virgo Cluster Catalog \cite[EVCC,][]{Kim2014}. VCC contains 2096 galaxies, of which 1277 are certain members, 574 are possible members, and 245 are background galaxies covering an area of $\sim$ 140 deg$^2$ \citep{Binggeli1985}. EVCC comprises 1589 galaxies, of which 913 belong to the original VCC with an addition of 676 new member galaxies, covering an area of $\sim$ ~725 deg$^2$. We have combined the EVCC galaxies and VCC galaxies to build a base sample of 2,772 galaxies. The analysis presented in this work is based on a sample of galaxies in the Virgo cluster ($175^\circ < \text{R.A.(J2000)} < 200^\circ$; $-4^\circ < \text{Dec.(J2000)} < 25^\circ$) extracted from the VCC and EVCC, with $m_B < 18$ mag, which marks the completeness limit. Assuming a Virgo cluster distance of 16.5 Mpc \citep{2007Mei}, this corresponds to an absolute magnitude limit of approximately $M_B \leq -13.085$ mag. This study focuses only on elliptical and lenticular galaxy populations, excluding S0a, dE, and dS0 galaxies. From the compiled list of galaxies, we segregated ETGs based on their membership and morphology, which left us with a total number of 109 galaxies, of which 43 are ellipticals, and 66 are lenticular galaxies. 

\par
%https://arxiv.org/pdf/0801.2113

We used the available data of a wide wavelength range from Galaxy Evolution Explorer \cite[GALEX]{Martin2005} in UV, Sloan Digital Sky Survey \cite[SDSS]{SDSS2000} in optical, the Two Micron All Sky Survey \cite[2MASS]{20062MASS} and Wide-field Infrared Survey Explorer \cite[WISE]{wrightwise2010} in IR. Varying apertures across surveys can impact SED analysis by affecting measured fluxes and introducing biases, making it essential to ensure consistency in the apertures used to avoid skewed results. We carefully selected datasets where integrated or total fluxes were derived from surface photometry isophotal profiles or asymptotic magnitudes, which minimize the effects of aperture variations. In this study, we utilized the public data archives NASA/IPAC Extragalactic Database \citep[NED]{NED1991} and HyperLeda \citep{Hyperleda2003} for data collection. For GALEX photometry, we used data from The GALEX Ultraviolet Atlas of Nearby Galaxies \citep{2007gildepaz} and The GALEX/S4G Surface Brightness and Color Profiles Catalog \citep{2018Bouquin_ph}, ensuring that integrated UV fluxes were computed from surface photometry profiles. For SDSS, we adopted MAG AUTO from The Extended Virgo Cluster Catalog \citep{Kim2014} as the total magnitude. Additionally, we obtained 2MASS data from the 2MASS Large Galaxy Atlas \citep{2003Jarrett} and The 2MASS Extended Sources Catalog \citep{2003Skrutskie}, which employed standard 2MASS aperture isophotal analysis and mean surface brightness to calculate integrated fluxes from J, H, and Ks mosaic images. For WISE photometry, we obtained total fluxes from the AllWISE Data Release \citep{2014Cutri,2017Cluver}. Through these careful selections, we ensured that the fluxes used in our analysis are consistent and minimally affected by aperture variations, thereby reinforcing the robustness of our SED analysis.\par

Hence, out of 109 galaxies, we found 106 galaxies with the photometric data for the range of wavelengths from Far -- UV (FUV) to Mid -- IR (MIR). The data obtained are corrected for Galactic extinction using \cite{Calzetti2000} extinction law. 
\section{Identification of Quiescent UV upturn galaxies}
\label{sec:identific}
Previous studies confirmed that ETGs are not all passively evolved, and $\sim$ 30\% show significant residual star formation activity in them \citep{Yi2005,Kaviraj2007,sheen2016}. Hence, identifying the red sequence galaxies is the first step in our study to comprehend the UV Upturn phenomenon. Since the predominant inhabitants are cool, low-mass stars either in their main sequence or later stages of evolution, red sequence galaxies emit mostly at longer wavelengths \citep{Renzini2006,O'Connell1999}. \par 

\begin{figure}
     \centering
     \includegraphics[width=\columnwidth]{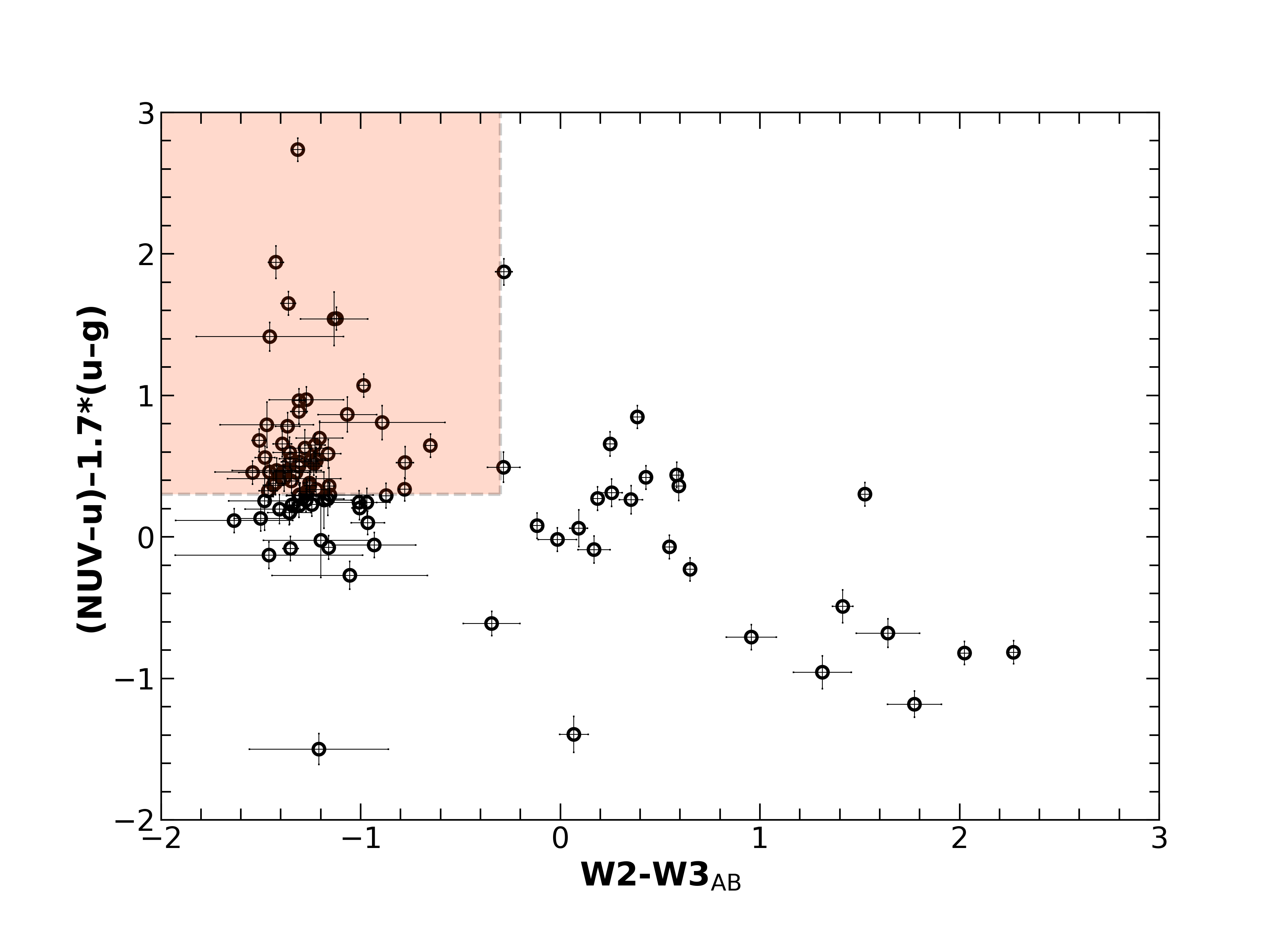}
     \caption{(NUV -- u) -- 1.7 * (u -- g) vs WISE (W2 -- W3) CCD for the galaxy sample. The red region on the plot indicates where the red sequence galaxies are located. Any galaxies situated outside this red region are classified as RSF galaxies.}
     \label{fig:wisecut}
 \end{figure}
Quiescent galaxies are expected to have a sharp decline in their flux from the {\it u} band to NUV. However, any residual star formation will cause a flat spectrum towards the shorter wavelengths. Thus, a UV--optical color such as (NUV -- r) or (NUV -- u) can be used to identify red sequence galaxies from those with any amount of residual star formation \citep{Schawinski2007,Hernandez-perez2014}. \cite{philips2020} has modified the color criteria from (NUV -- u) to (NUV -- u) -- 1.7 * (u -- g) for better identification of the red sequence galaxies. Constraining the selection criteria to a single color could also produce a sample of red sequence galaxies containing star-forming ones. Hence, \citet{philips2020} introduced an additional constraint based on the WISE color (W2 -- W3) to exclude galaxies exhibiting residual star formation. In this study, we employ the color criteria outlined by \cite{philips2020} to identify the red sequence galaxies. Figure \ref{fig:wisecut} illustrates the color-color diagram (CCD) of our initial sample. The shaded region indicates that the limit y > 0.3 and x < -0.3 is applied to eliminate galaxies with residual star formation. We separated our data sample into red sequence and residual star-forming (hereafter RSF) galaxies based on the selection criteria. Among the 106 ETGs in our initial sample, 49 galaxies are found to be red sequence galaxies (25 elliptical and 24 lenticular galaxies). The remaining 57 objects are considered as the RSF galaxies.

Galaxies exhibiting UV upturn will be present in our red sequence galaxy sample \citep{Ali2018a,philips2020}. FUV -- NUV vs NUV -- r (FUV -- NUV -- r) CCD is widely used for identifying the old population of quiescent galaxies showing UV upturn. \cite{Yi2005,Yi2011} have shown that the UV spectral slope becomes flat at FUV -- NUV = 0.9; hence,  FUV -- NUV < 0.9 represents the rising UV slope. Also, to account for the residual star formation activity, they suggested a NUV -- r > 5.4 as the limit. On the contrary, \cite{Smith2012} and \citet{philips2020} have suggested that the UV color needs to be more of a continuous range rather than discrete limits specified by \citet{Yi2011} and proposed a range for FUV -- NUV color between 0.7 and 2.1 and NUV -- r between 5.2 and 6.5 to identify the UV upturn region. Hence, we have considered both the criteria introduced by \citet{Yi2011} and \citet{philips2020} in this study.

\begin{figure}
      \centering
     \includegraphics[width=\columnwidth]{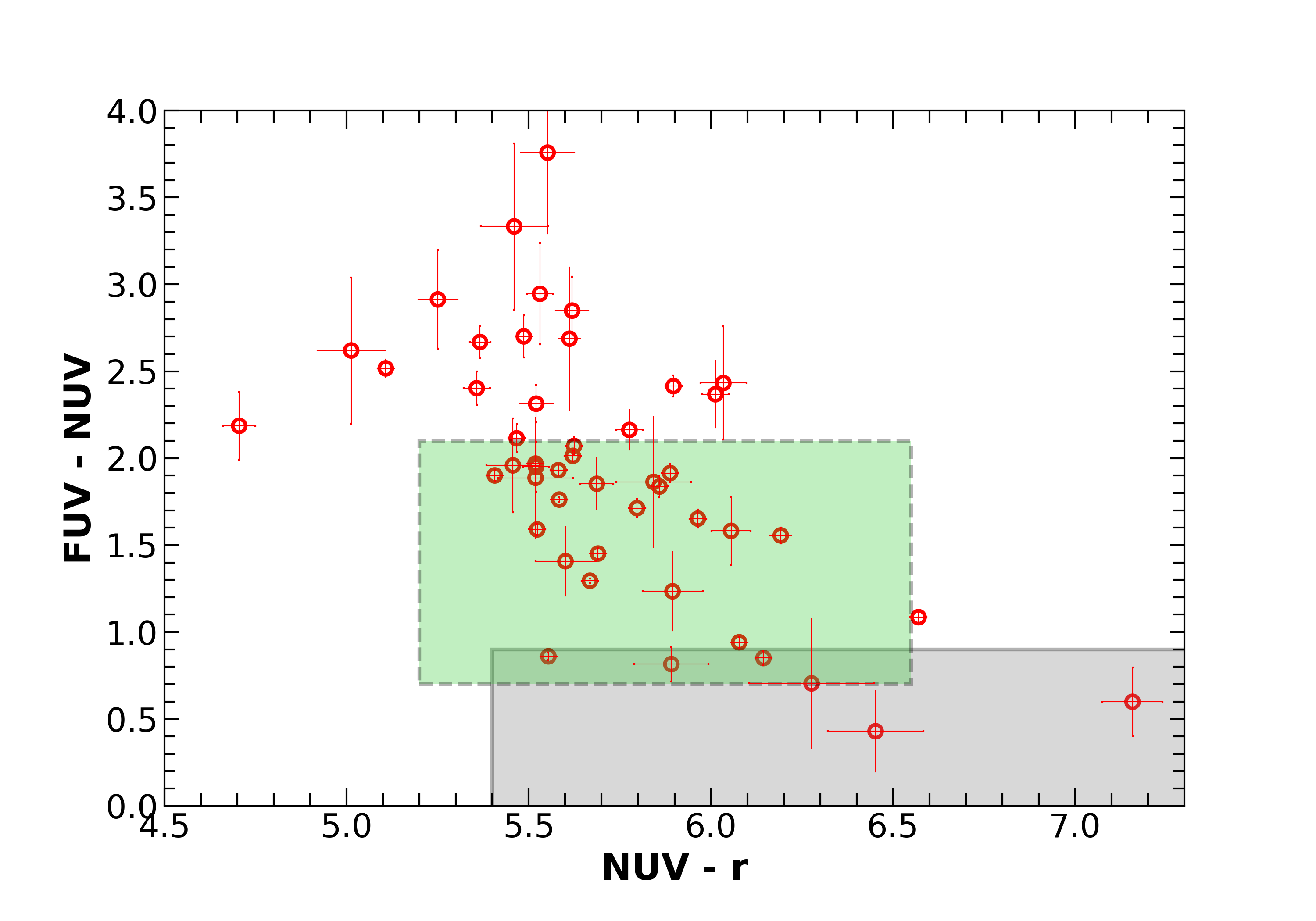}
     \caption{FUV -- NUV vs. NUV -- r CCD for the 49 red sequence galaxies of Virgo cluster. In the figure, the grey shaded region indicates the FUV -- NUV $>$ 0.9 and NUV -- r $>$ 5.4 limits obtained from \citet{Yi2011}. The green-colored region represents the continuous limits proposed by \citet{philips2020}.}
     \label{fig:FUVNUVCCD} 
\end{figure}

Figure \ref{fig:FUVNUVCCD} shows the FUV -- NUV -- r CCD plot for 49 ETGs. We show two selection criteria with different colored regions in the image. The grey-shaded region follows the criterion of \cite{Yi2011}, and the green-shaded region represents the limits proposed by \cite{philips2020}. We observed six galaxies (four elliptical and two lenticular) meeting the \cite{Yi2011} limit and 27 galaxies (17 elliptical and 10 lenticular) conforming to the \cite{philips2020} limit. Only four galaxies satisfied both criteria. Therefore, and altogether, we combined both scenarios and defined 29 galaxies as possible UV upturn galaxies.

\begin{table}
\caption{The filters and their wavelength used in the SED fitting.}
\label{tab:filters}
\begin{tabular}{lll|lll}
Telescope              & Filter & \textbf{$\lambda$ ($\AA$)} & Telescope & Filter & \textbf{$\lambda$ ($\AA$)} \\ \hline
\multirow{2}{*}{}{Galex} & FUV             & 1545 & \multirow{3}{*}{}{2MASS} & J               & 12350             \\  
                         & NUV             & 2344  &                     & H               & 16620             \\ 
\multirow{5}{*}{}{SDSS} &u               & 3572    &                       &K               & 21590             \\ 
                        & g               & 4750   &\multirow{4}{*}{}{WISE} & WISE 1               & 33526             \\        
                        & r               & 6204    &                       & WISE 2               & 46028             \\           
                        & i               & 7519   &                          & WISE 3               & 115608           \\           
                        & z               & 8992  &                             & WISE 4               & 220883            \\ \hline
\end{tabular}
\end{table}

\begin{figure}
      \centering
     \includegraphics[width=0.9\columnwidth]{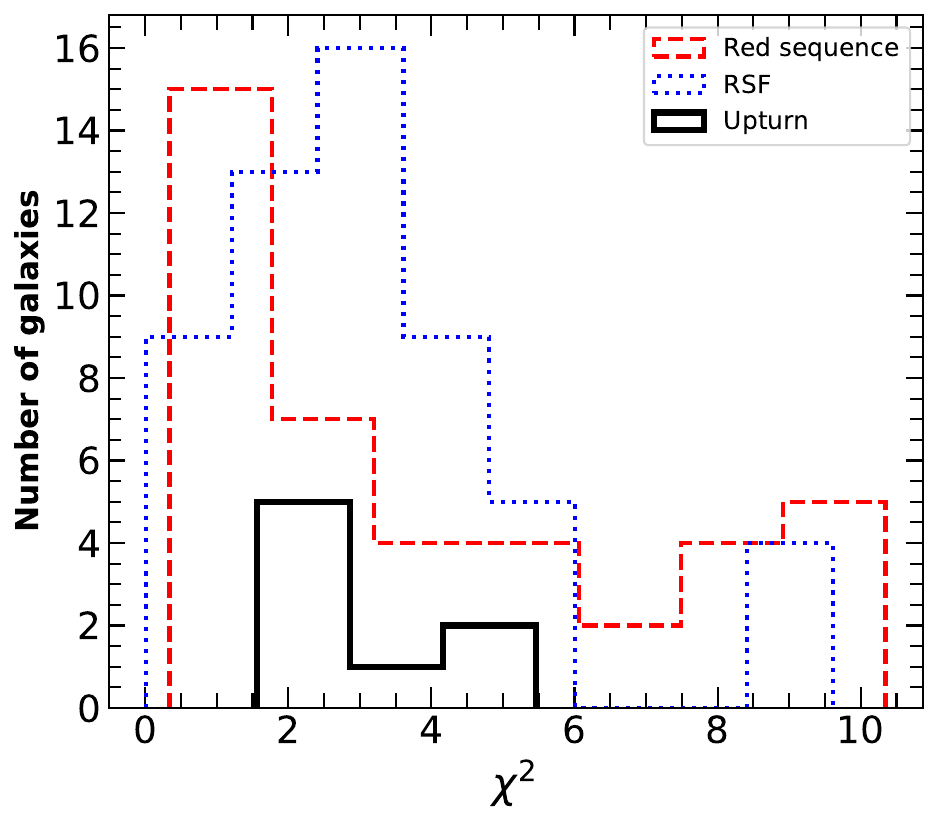}
     \caption{Histogram shows the reduced chi-square distribution of the best-fit SED models for our galaxy sample. The distributions for the red sequence, RSF, and upturn samples are shown with red-dashed, blue-dotted, and black-solid lines, respectively. }
     \label{fig:chisq} 
\end{figure}

\subsection{SEDs of UV upturn galaxies}

From the previous analysis, we have identified 29 galaxies as possible UV upturn galaxies. In this section, we develop the SED for all 106 ETGs, including 29 possible UV upturn galaxies. In this section, we will exploit an innovative method for investigating UV upturn galaxies, especially in cases where we lack spectroscopic data. This method revolves around SED modeling, a comprehensive technique that extracts complete information about a galaxy by fitting observed data onto theoretical models. This process entails selecting the most suitable models for each parameter. A detailed SED for a galaxy with complete UV to IR data can give access to information related to all components present in it. We use the SED fitting method using CIGALE software \citep{Boquien2019} to visualize and understand the UV upturn phenomenon in all the ETG samples. The filters utilized and their corresponding mean wavelengths are given in Table \ref{tab:filters}.

\begin{figure*}
    \centering
    \includegraphics[width=0.9\textwidth]{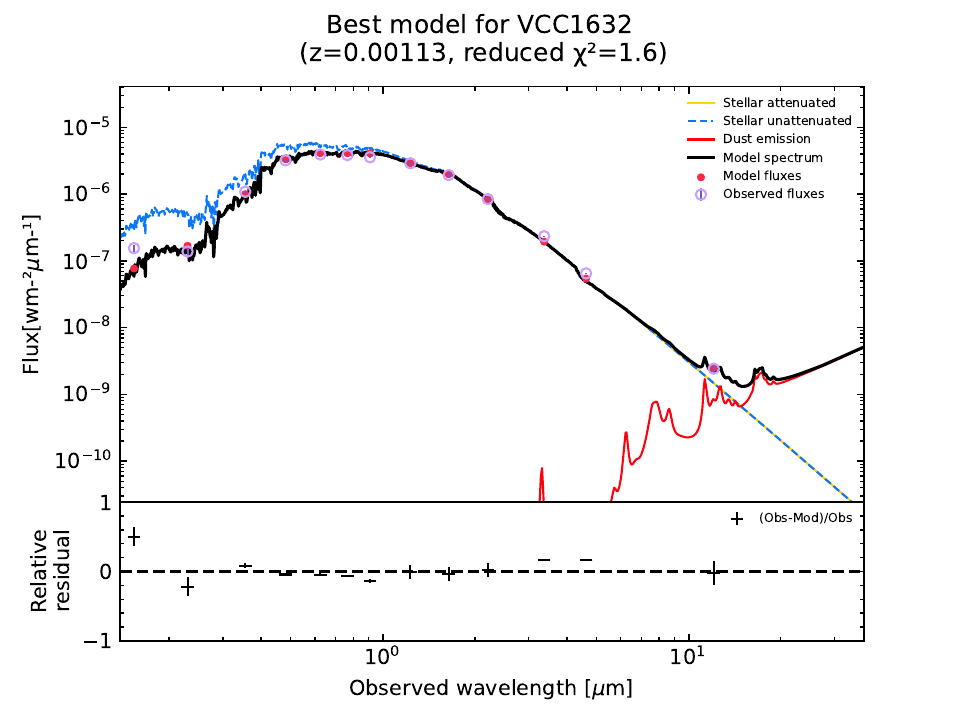}
    \caption{An illustration of UV upturn feature in the SED of galaxy VCC 1632, built using CIGALE. The best-fit model spectrum of the galaxy was obtained using the parameters (listed in Table\ref{tab:parameters}) shown in a black solid line. Blue circles indicate observed fluxes and fluxes extracted from the best model denoted by red points. }
    \label{fig:VCC1632}
\end{figure*}

We have carefully selected the models to compute the best-fit SED for each galaxy. Table \ref{tab:parameters} shows the models selected and the parameters chosen for computing every model in detail \citep{Burgarella2005, Boquien2019, Boquien2020}. CIGALE uses the compilation method of various parameters to form the best fit SED for a galaxy. Here, we build SEDs for all the ETGs using the parameters given in Table \ref{tab:parameters}. As galaxies evolve and interact over cosmic times, their star formation rates (SFR) exhibit varied patterns, ranging from intense bursts to quiescent phases. The young stellar populations in this study are defined as the energy contributions from O, B, and A-type stars with ages < 200 Myrs, typically modeled in the UV part of the spectrum in young galaxies. Here, we employ the \textit{'sfhdelayed'} model to elucidate the evolution of star formation history (SFH). The \textit{'sfhdelayed'} model describes the SFH of galaxies through a gradual increase in star formation rate from an onset age to a peak time, followed by a smooth decline. The model efficiently captures the SFH of ETGs, particularly when $\tau$ (the time at which the SFR peaks) is small \citep{Boquien2019}. Notably, the parameter SFH does not incorporate a burst; rather, it features an early smooth star formation, with the mass fraction of the late burst population set to zero.

The choice of stellar population synthesis models, including evolved stellar populations such as TP-AGB, HB, and PAGB, can indeed be critical for accurately modeling the UV part of the spectrum in old galaxies \citep{1998Maraston,2006AMaraston}. We utilized the \citet{Maraston2005} model in our analysis and fitting procedure to ensure that these critical stellar populations are appropriately accounted for. Considering the contributions from evolved stellar populations, we used metallicity values of 0.01, 0.02, and 0.04. Also, galaxies contain dust, which is highly effective at absorbing short-wavelength radiation. To account for this dust attenuation, we employed the \textit{`dustatt\_modified\_starburst'} module of CIGALE, based on the attenuation law proposed by \cite{Calzetti2000}. This model varies the color excess of the nebular lines (E($B-V$)$_{\text{lines}}$) and applies a reduction factor to compute the stellar continuum attenuation. The slope of the attenuation curve can be modified using a power-law function. Emission lines are attenuated using the Small Magellanic Cloud extinction law, with an $R_V$ value of 3.1 \citep{Boquien2019}. Additionally, we utilized the dust emission model by \citet{2014Draine} to model the dust emissions within the galaxies,  which builds on earlier work by \citet{Draine2007}.  This model provides empirical templates for dust emission, dividing the emission into components associated with different radiation field intensities. Key parameters include the mass fraction of polycyclic aromatic hydrocarbons (PAHs) and the minimum radiation field intensity ($u_{\text{min}}$). The flexibility of the \textit{`dl2014'} model allows for a detailed fit of the dust SED, accommodating a wide range of physical conditions and providing a nuanced understanding of the dust emission properties of galaxies. Utilizing the aforementioned models and the parameters outlined in Table \ref{tab:parameters}, we conducted SED fitting for all the ETGs included in this study.

\begin{figure*}
    \centering
    \includegraphics[width=\textwidth]{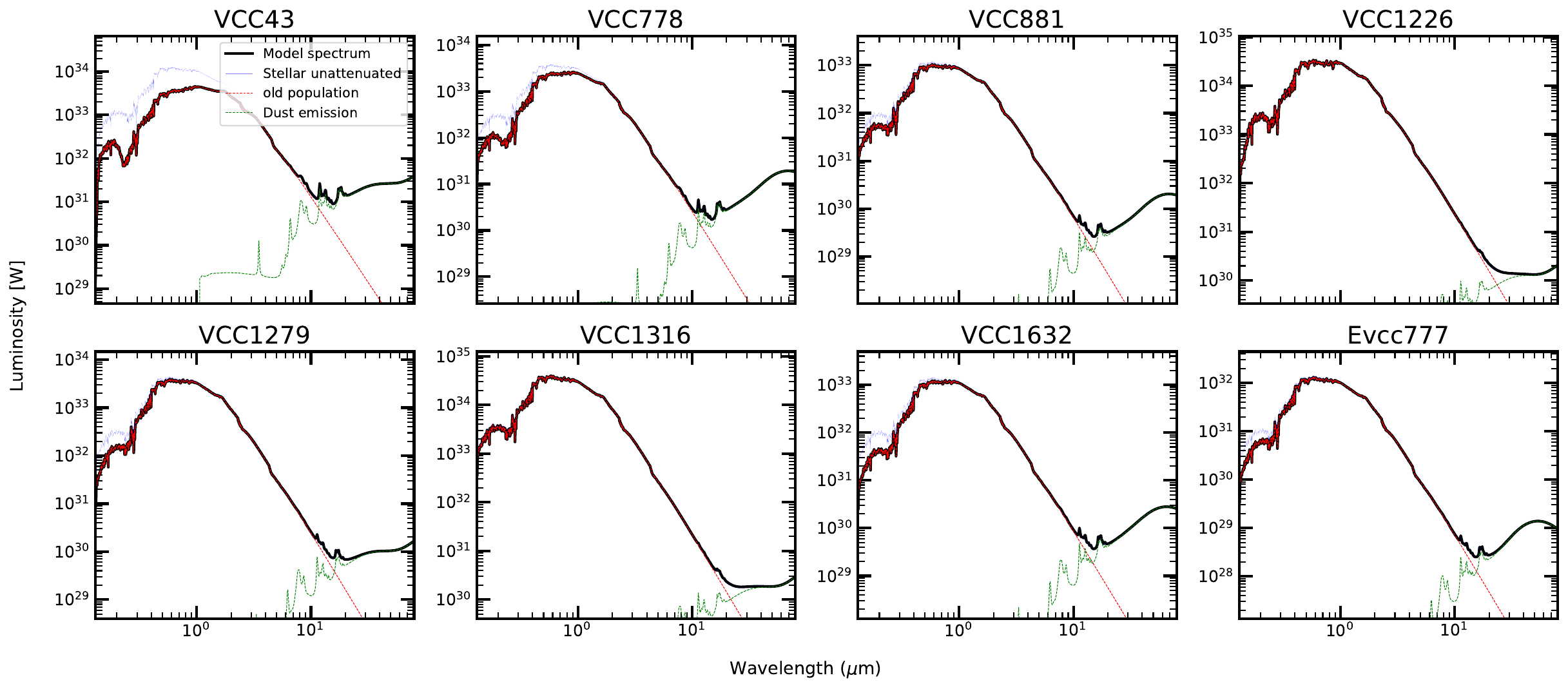}
    \caption{An illustration of the UV upturn feature in the spectral energy distribution of all eight galaxies built using CIGALE. The best-fit model spectrum of each galaxy was obtained using the parameters (listed in Table \ref{tab:parameters}) shown as a black solid line. The red dashed line depicts the attenuated old stellar population, while the blue line shows the unattenuated total spectrum. The green dotted lines are the dust emission.} 
    \label{fig:all_upt}
\end{figure*}

We performed a visual analysis of the SED of 29 potential upturn candidates. We identified eight galaxies exhibiting a distinctive rise between 1200 and 2500 \AA. Consequently, eight potential upturn galaxies with this discernible feature in the UV region of the SED were confirmed as UV upturn galaxies within the Virgo cluster (VCC 43, 778, 881, 1226, 1279, 1316, 1632, and EVCC 777). The SED fits for our galaxies are considered reliable, as we have evaluated them through an analysis of the overall chi-square distribution and a visual inspection. Figure \ref{fig:chisq} illustrates the histogram of the reduced chi-square values obtained from the SED fit for galaxies. As a result, using this SED building technique, the UV upturn phenomenon is visualized in these identified galaxies, and the galaxy properties are obtained. The best model fit SED obtained for the UV upturn galaxy VCC1632 is shown in Figure \ref{fig:VCC1632}, and the upturn feature is evident in the UV region of the SED. All best model fit SED for all other upturn galaxies are shown in Figure \ref{fig:all_upt}.

%The output parameters are given in Table \ref{tab:Cigale_output}
\begin{table*}
\caption{Models and parameter values used to build SED for ETGs. The values are derived from the SED fit and compared with the best available literature values \citep{Boquien2019,Hunt2019,Gao2018}. We use the default values for the parameters that are not listed here.}
\begin{tabular}{|c|c|c|}

\hline 
Module & Parameter & Value \\ \hline
\multirow{3}{*}{Star formation history (sfhdelayed)} & tau\_main & 0.1 - 3 Gyr \\ 
& age\_main & 10 - 13 Gyr \\ 
%& tau\_burst & 9, 10 Gyr \\ 
%& age\_burst & 9, 10 Gyr \\ 
& f\_burst & 0.0         \\\hline
\multirow{2}{*}{Stellar population (m2005)} & imf & 0 (salpeter) \\ 
& metallicity &0.01, 0.02, 0.04 \\ \hline
\multirow{3}{*}{Dust attenuation (dustatt\_modified\_starburst)} & E(B\_V) lines & 0.01 - 0.5 \\ 
& E(B\_V) factor & 0.01 - 0.9 \\ 
& powerlaw\_slope & -1, -0.1, -0.5, 0, 0.5, 1 \\ \hline

\multirow{4}{*}{Dust emission (dl2014)} & qpah & 0.47, 1.77, 3.90 \\
& umin & 0.1,1,5,10\\ \
& alpha & 1,2,3 \\ 
& gamma & 0.1,0.3,0.5,0.9\\ \hline
 
\end{tabular}
\label{tab:parameters}
\end{table*}

\begin{table*}
\caption{The SED fitting output values. The parameters are presented in the following order: Galaxy name, redshift, reduced chi square, star formation rate, e-folding time of the main stellar population, age of the main stellar population, and the logarithm of stellar mass.}
\label{tab:cig_out_tab}
\begin{tabular}{lllllll} \hline
Name    & redshift & $\chi^{2}$ & SFR ($M_\odot$$yr^{-1}$) & Tau main (Myr) & Age (Myr)     & Log $M^{*}$ ($M_\odot$) \\ \hline
VCC43   & 0.059    & 5.5        & 0.026 +/- 0.006          & 1640 +/- 890   & 12000 +/- 284 & 11.4 +/- 0.03           \\
VCC778  & 0.0045   & 2.0        & 0.005 +/- 0.001          & 1700 +/- 825   & 12000 +/- 473 & 10.68 +/- 0.05          \\
VCC881  & 0.0007   & 2.1        & 0.002 +/- 0.001          & 1800 +/- 801   & 12000 +/- 468 & 10.11 +/- 0.04          \\
VCC1226 & 0.003    & 2.2        & 0.002 +/- 0.001          & 1200 +/- 650    & 13000 +/- 478 & 11.8 +/- 0.05           \\
VCC1279 & 0.004    & 3.0        & 0.004 +/- 0.001          & 1510 +/- 930    & 12000 +/-476  & 10.56 +/- 0.03          \\
VCC1316 & 0.004    & 2.6        & 0.001 +/- 0.0009         & 1200 +/- 870   & 13000 +/- 478 & 11.69 +/- 0.02          \\
VCC1632 & 0.001    & 1.6        & 0.002 +/- 0.0004         & 1600 +/- 920   & 13000 +/- 477 & 10.32 +/- 0.03          \\
Evcc777 & 0.003    & 4.6        & 0.0001 +/- 0.00004       & 1900 +/- 810   & 13000 +/- 427 & 9.16 +/- 0.02      \\ \hline   
\end{tabular}
\end{table*}

%\subsection{Physical properties and CIGALE output parameters}
\begin{figure}
    \centering
    \includegraphics[width=0.9\columnwidth]{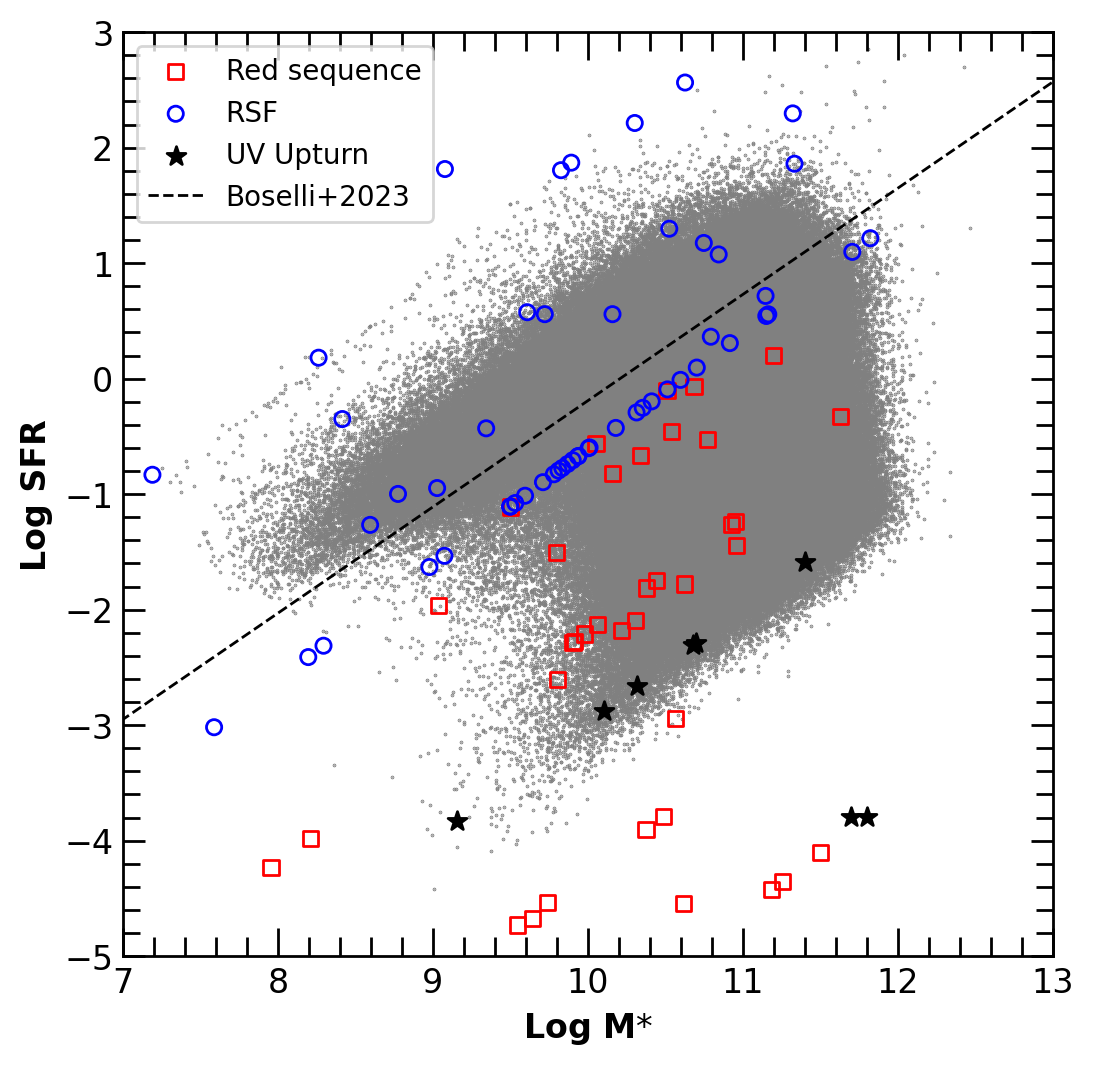}
    \caption{ Comparison between Red sequence, RSF, and UV upturn galaxies as a function of stellar mass and SFR. The blue circles, red squares, and black star markers represent the RSF, red sequence, and upturn galaxies respectively. The grey points are the galaxies obtained from the GSWLC \citep{2016Salim,2018Salim}. The black dashed line represents the Main Sequence relation for galaxies in the Virgo cluster with normal HI content, as obtained from \citep{2023BoselliMS}. }
    \label{fig:mstar_sfr}
\end{figure}

The UV upturn phenomenon is essentially characterized by UV emissions originating from evolved stellar evolutionary phases such as the HB and AGB phases, predominantly associated with low-mass, old stellar populations. Furthermore, the UV bump within the UV range ($\lambda$ < 3000$\AA$) is primarily attributed to the influence of old stellar populations. To understand the UV bump in all these galaxies from the old population, we present Figure \ref{fig:all_upt}. It is evident from the figure that all galaxies exhibit an upturn feature, with the fitted SED showing only contribution from the old stellar population. This suggests that, without using young stellar population models, we can fully model the UV contribution of upturn galaxies using CIGALE. In either case, this observation supports the notion that the UV emission in these galaxy types primarily originates from old stellar populations. Additionally, We compared the red sequence galaxies, RSF, and UV upturn galaxies based on the SED parameters: stellar mass ($M^{*}$) and star formation rate (SFR). Figure \ref{fig:mstar_sfr} illustrates the SFR-$M^{*}$ correlation among red, RSF, and upturn galaxies. Notably, all upturn galaxies display markedly low SFR.

Among the confirmed upturn galaxies in this study, VCC 1632, VCC 1316, VCC 881, and VCC 1226 have previously been examined as upturn galaxies. VCC 1632 (NGC 4552, M89) is a well-studied galaxy exhibiting upturn features, and \citet{Burstein1988} was the first to observe the upturn phenomenon in this galaxy using IUE spectra. \citet{Yi2005,Yi2011} suggested that the NUV flux of VCC 1632 is an empirical upper limit that can be produced by an old stellar population. Also, \citet{2022Bosellim89,2023Bosellim89} observed the H$\alpha$ emission peaks at the center of VCC 1632, primarily attributed to a diffuse ionized gas filament structure. They mentioned that the ionizing radiation may be generated by cooling flows or shocks. Similar H$\alpha$ emission was observed in VCC 1316 (NGC 4486, M87), where the ionizing radiation may also result from cooling flows or shocks \citep{2022Bosellim89,2023Bosellim89,2024Edler}. VCC 1316 was also observed for its upturn feature by \citet{Burstein1988} and is identified as a LINER galaxy \citep{2000Owen,2006VCC1316AGN,2010Boselli}.
\cite{1998ohl} noted that synchrotron emission contributes only about 17\% of the FUV light within a 20-arcsecond region. Furthermore, studies by \cite{2006Sohn} and \citet{2018Goudfrooij} indicate that UV emissions in M87 are primarily influenced by hot stellar populations in globular clusters. Therefore, it is important to recognize in this study that the observed SED is primarily due to these hot stellar populations, with a minor contribution from synchrotron emission. Additionally, \citet{2018Anderson} highlighted that the FUV continuum observed in this galaxy is associated with emission from evolved stars. \citet{2008salome_combes} also reported that there is no evidence of star formation within M87. VCC 881 (NGC 4406, M86) and VCC 1226 (NGC 4472, M49) were identified as upturn galaxies based on IUE spectra \citep{Burstein1988}. These galaxies have also been utilized in studying the UV flux emanating from old stellar populations \citep{Dorman1995,Dorman2003}. Hence, VCC 43 (NGC 4164), VCC 778 (NGC 4377), VCC 1279 (NGC 4478), and EVCC777 (IC 3436) have been identified as new upturn candidates in the Virgo Cluster region based on this study.
%However, it appears that AGN activity only plays a minor role in the far-UV emission of systems with very bright nuclei \citep{1998ohl,dantas2020}.
\subsection{Temperature and Metallicity}
\label{sec:bbody_mg2}
To understand more about the old stellar populations and their contribution to the upturn phenomena, we studied the temperature and metallicity properties of these galaxies. As mentioned in the introduction, the question of which stellar population contributes to the observed UV upturn phenomenon in ETGs remains a topic of debate. Using the confirmed sample of 8 upturn galaxies obtained from our SED analysis, we can better understand the origin of the UV upturn phenomena. Estimating the temperature in the UV region of these galaxies can provide valuable insights into the stellar population responsible for the phenomenon. Therefore, it's essential to consider higher temperature components within the galaxies, contributing to the UV excess and the conventional old stellar population. Here, we used the same method as \cite{Ali2018a} to estimate the temperature of the hot population contributing to the UV upturn phenomena. To model the old stellar population within the galaxy, we utilized the C09 models \citep{Conroy2009A}, generated through the FSPS software \citep{danforemanmackey2014}. For the hot stellar component, we employed theoretical blackbody models ranging from 1000 K to 40000 K, which were sourced from a virtual observatory SED analyzer (VOSA). Each blackbody was subjected to an extinction correction, varying the Av values from 0 to 1 in 0.1 intervals, applying the \citep{Calzetti2000} extinction curve. The blackbodies are also normalized to the C09 model in the NUV band, with a normalization parameter ranging from 1 percent to 500 percent in 50 percent intervals. When this parameter reaches 100\%, the blackbody flux matches that of the C09 model in the NUV band. We then fit the UV region of our UV upturn galaxies to these normalized blackbodies, determining the best fit through the least squares fit method \citep{Ali2018a}. The resulting extinction values are listed in Table \ref{tab:temp} under column \textit{Extinction}. This approach was adopted to prevent bias in our best-fit black body model, which could occur if adopting pre-reported extinction values for the sample galaxies. \par

\begin{figure}
    \centering
    \includegraphics[width=0.9\columnwidth]{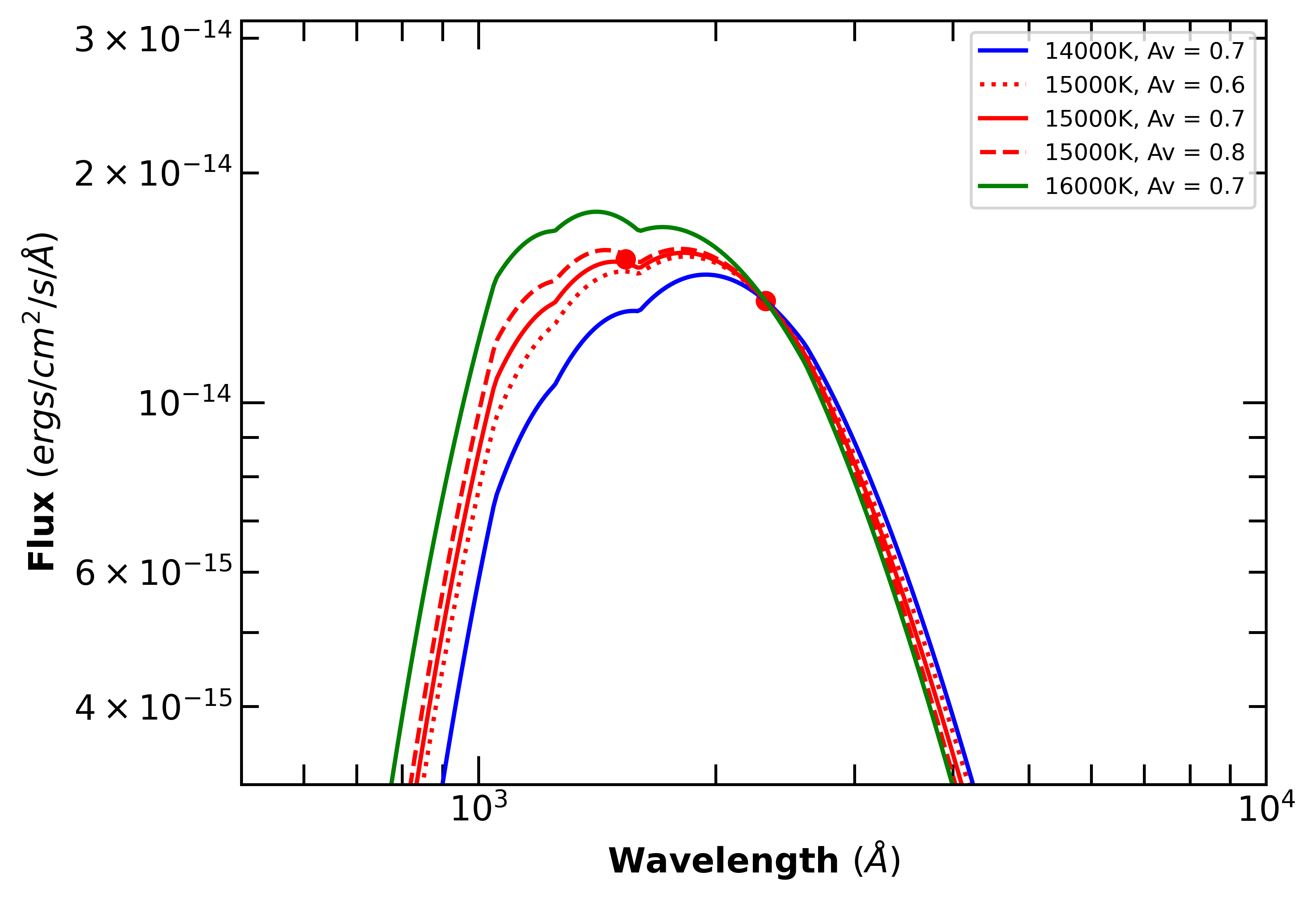}
    \caption{Shows the best-fitted blackbody temperature for the upturn galaxy VCC 1632. The blue, red, and green lines represent different blackbody temperatures: 14000K, 15000K, and 16000K, respectively. The dotted red lines depict the temperature of 15000K with varying Av values. The red continuous line represents the best-fitted blackbody for NGC 1632.}
    \label{fig:bbfit}
\end{figure}

Figure \ref{fig:bbfit} depicts the best-fitted blackbody curve for the galaxy VCC 1632. The dotted and dashed red lines in the figure represent the varying Av values. Our analysis reveals that a temperature of 15000K best fits the UV flux of galaxy VCC 1632. The best-fitted blackbody for all other galaxies is shown in Figure \ref{fig:BBfit_all}.
Hence, the temperature of the population contributing to the upturn feature in the 8 UV upturn confirmed galaxies is estimated. The result obtained from the analysis is given in Table \ref{tab:temp}. It is found that the temperature of the stellar population lies in the range of 14,000 K to 17,000 K. In quiescent ETGs, these temperatures can be produced only by low-mass stellar components in their evolved stages with helium-burning cores and hydrogen-burning shells as in HB stars or their progeny \citep{O'Connell1999,Brown2000}. According to this result, the confirmed UV upturn galaxies in the Virgo cluster were found to have a population of HB stars that are Helium-enriched.

\begin{table}
\caption{Table showing the temperature and derived parameters of UV upturn galaxies at the UV region from the model.}
\centering
\begin{tabular}{|c|c|c|c|c|}
\hline
\textbf{Galaxy} & \textbf{Temperature (K)} & \textbf{Extinction} & \textbf{Norm $\%$} \\ \hline
VCC43 & 18000 & 0.1  & 1 \\ 
VCC778 & 15000 & 0.9  & 51 \\ 
VCC881 & 13000 & 0.4  & 101  \\ 
VCC1226 & 13000 & 0.1  & 401 \\
VCC1279 & 14000 & 0.7  & 301 \\
VCC1316 & 14000  & 0.7  & 301\\
VCC1632 & 15000  & 0.7   & 401   \\ 
EVCC777  & 16000 & 0.1   & 151\\ 

\hline
\end{tabular}
\label{tab:temp}
\end{table}

\begin{figure}
    \centering
    \includegraphics[width=0.9\columnwidth]{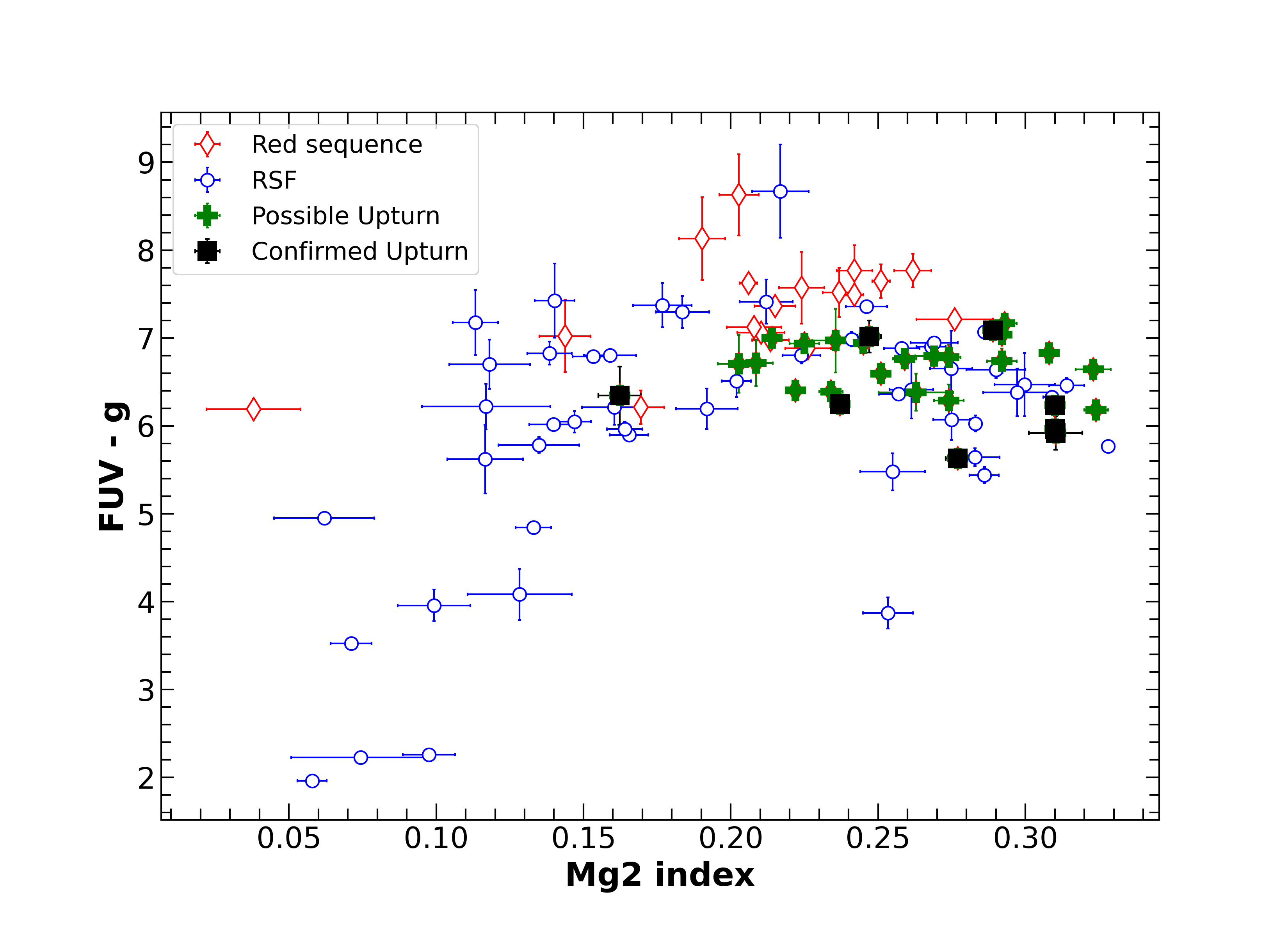}
    \caption{shows the relationship between the Mg2 Lick index and (FUV -- g) color for the galaxy sample. A trend is visible between metallicity and UV upturn, with the upturn sample showing a higher Mg2 index. Here, the RSF galaxies are marked in blue, and those red sequence galaxies are marked in red. All the possible and confirmed UV upturn galaxies are marked in green plus symbols and black-filled boxes, respectively.}
    \label{fig:Mg2Fuv}
\end{figure}
Furthermore, Mg2 strength is extensively used in previous studies to understand the age and metallicity properties of galaxies having old stellar populations \citep[][reference therein]{Almeida2012}. We utilized the Hyperleda database\footnote{http://atlas.obs-hp.fr/hyperleda/} to obtain the Mg2 strength for the sample galaxies. \citet{Boselli2005}, and \citet{Smith2012} studied the relation between UV upturn with age/metallicity and identified a correlation between (1550 -- V) with Mg2. They found that the upturn galaxies have higher Mg2 strength, while SFGs show a lower Mg2 strength. Also, \cite{Renzini2006} suggested that the helium enhancement due to very old stellar populations ($\gid$ 10 Gyr) exhibits strong Mg2 features in a galaxies' spectrum. A higher Mg2 index was observed for galaxies showing upturn, inferring their higher metallicity. This can be evidence of helium-enriched stellar populations in their evolved stages, which are assumed to be the sources of UV upturn. Using  FUV -- optical color against Mg2 plot, we intend to determine if any correlation exists between red sequence, RSF, and UV upturn galaxies. Figure \ref{fig:Mg2Fuv} shows the Mg2 index relation of our sample galaxies against UV -- optical color. Our analysis found that the UV upturn galaxies in the Virgo cluster region and Mg2 index relation follow the same trend as reported in the previous studies \citep{Boselli2005,Thomas2005,Smith2012}.

\section{Discussion}
\label{sec:discuss}
Studying the UV upturn in galaxies is crucial as it challenges our understanding of the star formation history and evolution of galaxies, especially ETGs, which are thought to have stopped forming stars long ago. James Webb Space Telescope (JWST) and Future space telescopes and observatories, such as the Indian Spectroscopic and Imaging Space Telescope (INSIST) and The Ultraviolet Explorer (UVEX), will provide unprecedented sensitivity and capabilities in the UV and near-infrared parts of the spectrum. These missions will significantly enhance our ability to study the UV upturn in galaxies, allowing us to probe fainter and more distant galaxies and refine our understanding of this phenomenon.
This study identified a confirmed set of UV upturn galaxies in the Virgo cluster environment. Our analysis indicates that the two FUV -- NUV -- r CCD limits proposed by \cite{Yi2011} and \cite{philips2020} successfully identify all potential upturn galaxies. Moreover, the limit proposed by \cite{Yi2011} is particularly significant since all six galaxies it selects have confirmed upturn properties in their SEDs. Based on a combined analysis using the well-defined UV--optical--IR color criteria and the new approach of SED fitting,  we identified eight confirmed UV upturn galaxies in the Virgo cluster. As this phenomenon concerns quiescent ETGs, their SED needs a typical early-type structure and an additional bump in the UV region, characterizing upturn. Using CIGALE software, we built SEDs with an increasing flux visible short-wards from 3000 \AA. SEDs of all identified upturn galaxies have a primary peak at the IR region and a clear secondary bump at the FUV region. 

A detailed analysis of all the obtained SEDs and output parameters was conducted for all eight confirmed upturn galaxies. As a reference, the SED of galaxy VCC1632, one of the confirmed UV upturn galaxies obtained through CIGALE, is shown in Figure \ref{fig:VCC1632}. %given in Table \ref{tab:Cigale_output}. 
The confirmed upturn galaxies show excess UV flux despite having low SFR rates when compared to normal ETGs having residual star formation or spiral galaxies with active star formation \cite[20 to 100 $M_\odot\ yr^{-1}$ ]{1998Kennicutt}. Also, the SFR of UV upturn galaxies is significantly lower compared to those of galaxies in the red sequence and RSF galaxies. All these galaxies have ages of their main stellar population from 12 Gyrs to 13 Gyrs, which relates to the age of low-mass stars in the main sequence. The output age parameter of CIGALE and primary SED characteristics clearly describes the dominance of old stellar populations in the galaxy. The UV bump at the FUV region indicates a low-mass stellar population in their evolved stages with high temperatures directed towards HB or AGB stars. This study indicates that, in UV upturn galaxies, we can fully model the UV contribution using CIGALE without relying on young stellar population models. The contributions from the old stellar population are sufficient to model the energy emission in the upturn galaxies, as clearly observed in the SEDs shown in Figure\ref{fig:all_upt}. To hold up the scenario of UV emission from the old stellar population, we have estimated the temperature of the hot population in upturn galaxies.

As explained in section \ref{sec:bbody_mg2}, a blackbody is fitted to the UV part of the C09 model, and the temperature of UV upturn galaxies is deduced and found to be in the range of 13,000 to 18,000 K. In quiescent ETGs, these temperatures can be produced only by hot low, mass stellar components in their evolved stages with helium-burning cores and hydrogen-burning shells as in EHB stars or their progeny \citep{Brown2000,O'Connell1999}. The temperature at the HB stage of stars depends on the mass-loss rates at the red giant branch (RGB) and the chemical composition, which in turn constitutes the metallicity and helium abundance \citep{Dcruz1996}. When metallicity increases, the temperature of HB stars reduces. However, when helium abundance increases, HB stars become hotter. Thus, the metallicity and helium abundance ratio determines the temperature at the HB stage \citep{Catelan2009}. In figure \ref{fig:Mg2Fuv}, upturn galaxies can be observed with higher metallicities. Their blackbody temperatures and spectroscopic properties, like the Mg2 index, direct towards the helium-rich EHB stellar population present in these galaxies.

\section{Summary}
\label{sec:summaty}
This study focused on a comprehensive sample of ETGs within the Virgo cluster region. Using the color limits obtained from \cite{Yi1997} and \cite{philips2020}, 29 possible UV upturn candidates are identified. Out of 29 possible candidates, 8 UV upturn-confirmed galaxies were detected from our SED analysis using CIGALE. We found out that the method of SED building is an effective tool for visualizing and confirming the UV upturn phenomenon.  All six galaxies obtained using the limits of \cite{Yi2011} explicitly show the UV upturn phenomenon in the SED. This suggests that the limit proposed by \cite{Yi2011} is more effective in identifying galaxies with the UV upturn characteristic than the limit set by \cite{philips2020}. In this study, we modeled the SED of UV upturn galaxies in the Virgo cluster using CIGALE. The best-fit parameters indicate that these galaxies are ETGs that significantly exhibit UV emission. Our findings suggest that this UV contribution originates from old stellar populations, with evolved stars playing a significant role. This result was achieved without invoking a burst model, demonstrating the ability to model the UV upturn in ETGs using these established stellar population models. With the stellar properties obtained from CIGALE, we observed a higher similarity between the upturn galaxies and the red sequence galaxies.  Also, the temperature of the stellar population contributing to the UV region is estimated to range between 13,000 K to 18,000 K. The temperature estimates and Mg2 strength support the consideration of contribution by low mass evolved stellar population as the origin of UV upturn phenomena.

\section{Acknowledgement}
We express our sincere gratitude to the referee for their valuable comments, which have significantly enriched the scientific content of the paper. AKR and SSK acknowledge the financial support from CHRIST (Deemed to be University, Bangalore) through the SEED money project (No: SMSS-2220, 12/2022). NK is supported by the Taiwan National Science and Technology Council and the Taiwan Ministry of Education. UK acknowledges the Department of Science and Technology (DST) for the INSPIRE FELLOWSHIP (IF180855). We sincerely thank Denis Burgarella and Pooja Sharma for their invaluable help and support during the CIGALE analysis. We express sincere gratitude to Sadman Ali for his help and support during the work. We thank Blesson Mathew, Koshy George, and Arun Roy for useful discussions on this work.  We thank the Center for Research, CHRIST (Deemed to be University), for all their support during this work.  This publication uses data from surveys, including GALEX, SDSS, WISE, and 2MASS. We gratefully thank all the individuals involved in the various teams for supporting those projects from the early stages of the design to launch and observations with it in orbit.
\section*{Data Availability}
The data underlying this article will be shared on reasonable request to the corresponding author.

%%%%%%%%%%%%%%%%%%%%%%%%%%%%%%%%%%%%%%%%%%%%%%%%%%



\bibliographystyle{mnras}
\bibliography{example} % if your bibtex file is called example.bib

\begin{thebibliography}{}
\makeatletter
\relax
\def\mn@urlcharsother{\let\do\@makeother \do\$\do\&\do\#\do\^\do\_\do\%\do\~}
\def\mn@doi{\begingroup\mn@urlcharsother \@ifnextchar [ {\mn@doi@} {\mn@doi@[]}}
\def\mn@doi@[#1]#2{\def\@tempa{#1}\ifx\@tempa\@empty \href {http://dx.doi.org/#2} {doi:#2}\else \href {http://dx.doi.org/#2} {#1}\fi \endgroup}
\def\mn@eprint#1#2{\mn@eprint@#1:#2::\@nil}
\def\mn@eprint@arXiv#1{\href {http://arxiv.org/abs/#1} {{\tt arXiv:#1}}}
\def\mn@eprint@dblp#1{\href {http://dblp.uni-trier.de/rec/bibtex/#1.xml} {dblp:#1}}
\def\mn@eprint@#1:#2:#3:#4\@nil{\def\@tempa {#1}\def\@tempb {#2}\def\@tempc {#3}\ifx \@tempc \@empty \let \@tempc \@tempb \let \@tempb \@tempa \fi \ifx \@tempb \@empty \def\@tempb {arXiv}\fi \@ifundefined {mn@eprint@\@tempb}{\@tempb:\@tempc}{\expandafter \expandafter \csname mn@eprint@\@tempb\endcsname \expandafter{\@tempc}}}

\bibitem[\protect\citeauthoryear{{Ali}, {Bremer}, {Phillipps}  \& {De Propris}}{{Ali} et~al.}{2018}]{Ali2018a}
{Ali} S.~S.,  {Bremer} M.~N.,  {Phillipps} S.,   {De Propris} R.,  2018, \mn@doi [\mnras] {10.1093/mnras/sty227}, \href {https://ui.adsabs.harvard.edu/abs/2018MNRAS.476.1010A} {476, 1010}

\bibitem[\protect\citeauthoryear{{Anderson} \& {Sunyaev}}{{Anderson} \& {Sunyaev}}{2018}]{2018Anderson}
{Anderson} M.~E.,  {Sunyaev} R.,  2018, \mn@doi [\aap] {10.1051/0004-6361/201732510}, \href {https://ui.adsabs.harvard.edu/abs/2018A&A...617A.123A} {617, A123}

\bibitem[\protect\citeauthoryear{{Bertola}, {Capaccioli}, {Holm}  \& {Oke}}{{Bertola} et~al.}{1980}]{Bertola1980}
{Bertola} F.,  {Capaccioli} M.,  {Holm} A.~V.,   {Oke} J.~B.,  1980, \mn@doi [\apjl] {10.1086/183236}, \href {https://ui.adsabs.harvard.edu/abs/1980ApJ...237L..65B} {237, L65}

\bibitem[\protect\citeauthoryear{{Binggeli}, {Sandage}  \& {Tammann}}{{Binggeli} et~al.}{1985}]{Binggeli1985}
{Binggeli} B.,  {Sandage} A.,   {Tammann} G.~A.,  1985, \mn@doi [\aj] {10.1086/113874}, \href {https://ui.adsabs.harvard.edu/abs/1985AJ.....90.1681B} {90, 1681}

\bibitem[\protect\citeauthoryear{{Boquien}}{{Boquien}}{2020}]{Boquien2020}
{Boquien} M.,  2020, in American Astronomical Society Meeting Abstracts \#235. p. 228.01

\bibitem[\protect\citeauthoryear{{Boquien}, {Burgarella}, {Roehlly}, {Buat}, {Ciesla}, {Corre}, {Inoue}  \& {Salas}}{{Boquien} et~al.}{2019}]{Boquien2019}
{Boquien} M.,  {Burgarella} D.,  {Roehlly} Y.,  {Buat} V.,  {Ciesla} L.,  {Corre} D.,  {Inoue} A.~K.,   {Salas} H.,  2019, \mn@doi [\aap] {10.1051/0004-6361/201834156}, \href {https://ui.adsabs.harvard.edu/abs/2019A&A...622A.103B} {622, A103}

\bibitem[\protect\citeauthoryear{{Boselli} et~al.,}{{Boselli} et~al.}{2005}]{Boselli2005}
{Boselli} A.,  et~al., 2005, \mn@doi [\apjl] {10.1086/444534}, \href {https://ui.adsabs.harvard.edu/abs/2005ApJ...629L..29B} {629, L29}

\bibitem[\protect\citeauthoryear{{Boselli} et~al.,}{{Boselli} et~al.}{2010}]{2010Boselli}
{Boselli} A.,  et~al., 2010, \mn@doi [\aap] {10.1051/0004-6361/201014534}, \href {https://ui.adsabs.harvard.edu/abs/2010A&A...518L..61B} {518, L61}

\bibitem[\protect\citeauthoryear{{Boselli} et~al.,}{{Boselli} et~al.}{2011}]{boselli2011}
{Boselli} A.,  et~al., 2011, \mn@doi [\aap] {10.1051/0004-6361/201016389}, \href {https://ui.adsabs.harvard.edu/abs/2011A&A...528A.107B} {528, A107}

\bibitem[\protect\citeauthoryear{{Boselli} et~al.,}{{Boselli} et~al.}{2018}]{2018BoselliVestige}
{Boselli} A.,  et~al., 2018, \mn@doi [\aap] {10.1051/0004-6361/201732407}, \href {https://ui.adsabs.harvard.edu/abs/2018A&A...614A..56B} {614, A56}

\bibitem[\protect\citeauthoryear{{Boselli} et~al.,}{{Boselli} et~al.}{2022}]{2022Bosellim89}
{Boselli} A.,  et~al., 2022, \mn@doi [\aap] {10.1051/0004-6361/202142482}, \href {https://ui.adsabs.harvard.edu/abs/2022A&A...659A..46B} {659, A46}

\bibitem[\protect\citeauthoryear{{Boselli} et~al.,}{{Boselli} et~al.}{2023a}]{2023BoselliMS}
{Boselli} A.,  et~al., 2023a, \mn@doi [\aap] {10.1051/0004-6361/202244267}, \href {https://ui.adsabs.harvard.edu/abs/2023A&A...669A..73B} {669, A73}

\bibitem[\protect\citeauthoryear{{Boselli} et~al.,}{{Boselli} et~al.}{2023b}]{2023Bosellim89}
{Boselli} A.,  et~al., 2023b, \mn@doi [\aap] {10.1051/0004-6361/202346506}, \href {https://ui.adsabs.harvard.edu/abs/2023A&A...675A.123B} {675, A123}

\bibitem[\protect\citeauthoryear{{Bouquin} et~al.,}{{Bouquin} et~al.}{2018}]{2018Bouquin_ph}
{Bouquin} A. Y.~K.,  et~al., 2018, \mn@doi [\apjs] {10.3847/1538-4365/aaa384}, \href {https://ui.adsabs.harvard.edu/abs/2018ApJS..234...18B} {234, 18}

\bibitem[\protect\citeauthoryear{{Brown}, {Ferguson}, {Stanford}  \& {Deharveng}}{{Brown} et~al.}{1998}]{Brown1998}
{Brown} T.~M.,  {Ferguson} H.~C.,  {Stanford} S.~A.,   {Deharveng} J.-M.,  1998, \mn@doi [\apj] {10.1086/306079}, \href {https://ui.adsabs.harvard.edu/abs/1998ApJ...504..113B} {504, 113}

\bibitem[\protect\citeauthoryear{{Brown}, {Bowers}, {Kimble}, {Sweigart}  \& {Ferguson}}{{Brown} et~al.}{2000}]{Brown2000}
{Brown} T.~M.,  {Bowers} C.~W.,  {Kimble} R.~A.,  {Sweigart} A.~V.,   {Ferguson} H.~C.,  2000, \mn@doi [\apj] {10.1086/308566}, \href {https://ui.adsabs.harvard.edu/abs/2000ApJ...532..308B} {532, 308}

\bibitem[\protect\citeauthoryear{{Burgarella}, {Buat}  \& {Iglesias-P{\'a}ramo}}{{Burgarella} et~al.}{2005}]{Burgarella2005}
{Burgarella} D.,  {Buat} V.,   {Iglesias-P{\'a}ramo} J.,  2005, \mn@doi [\mnras] {10.1111/j.1365-2966.2005.09131.x}, \href {https://ui.adsabs.harvard.edu/abs/2005MNRAS.360.1413B} {360, 1413}

\bibitem[\protect\citeauthoryear{{Burstein}, {Bertola}, {Buson}, {Faber}  \& {Lauer}}{{Burstein} et~al.}{1988}]{Burstein1988}
{Burstein} D.,  {Bertola} F.,  {Buson} L.~M.,  {Faber} S.~M.,   {Lauer} T.~R.,  1988, \mn@doi [\apj] {10.1086/166304}, \href {https://ui.adsabs.harvard.edu/abs/1988ApJ...328..440B} {328, 440}

\bibitem[\protect\citeauthoryear{{Calzetti}, {Armus}, {Bohlin}, {Kinney}, {Koornneef}  \& {Storchi-Bergmann}}{{Calzetti} et~al.}{2000}]{Calzetti2000}
{Calzetti} D.,  {Armus} L.,  {Bohlin} R.~C.,  {Kinney} A.~L.,  {Koornneef} J.,   {Storchi-Bergmann} T.,  2000, \mn@doi [\apj] {10.1086/308692}, \href {https://ui.adsabs.harvard.edu/abs/2000ApJ...533..682C} {533, 682}

\bibitem[\protect\citeauthoryear{{Catelan}}{{Catelan}}{2009}]{Catelan2009}
{Catelan} M.,  2009, Astrophysics and Space Science Proceedings, \href {https://ui.adsabs.harvard.edu/abs/2009ASSP....7..175C} {7, 175}

\bibitem[\protect\citeauthoryear{{Cluver}, {Jarrett}, {Dale}, {Smith}, {August}  \& {Brown}}{{Cluver} et~al.}{2017}]{2017Cluver}
{Cluver} M.~E.,  {Jarrett} T.~H.,  {Dale} D.~A.,  {Smith} J. D.~T.,  {August} T.,   {Brown} M.~J.~I.,  2017, \mn@doi [\apj] {10.3847/1538-4357/aa92c7}, \href {https://ui.adsabs.harvard.edu/abs/2017ApJ...850...68C} {850, 68}

\bibitem[\protect\citeauthoryear{{Code} \& {Welch}}{{Code} \& {Welch}}{1979}]{Code1979}
{Code} A.~D.,  {Welch} G.~A.,  1979, \mn@doi [\apj] {10.1086/156825}, \href {http://adsabs.harvard.edu/abs/1979ApJ...228...95C} {228, 95}

\bibitem[\protect\citeauthoryear{{Conroy} \& {Gunn}}{{Conroy} \& {Gunn}}{2010}]{Conroy2009A}
{Conroy} C.,  {Gunn} J.~E.,  2010, \mn@doi [\apj] {10.1088/0004-637X/712/2/833}, \href {https://ui.adsabs.harvard.edu/abs/2010ApJ...712..833C} {712, 833}

\bibitem[\protect\citeauthoryear{{Conroy}, {Gunn}  \& {White}}{{Conroy} et~al.}{2009}]{Conroy2009}
{Conroy} C.,  {Gunn} J.~E.,   {White} M.,  2009, \mn@doi [\apj] {10.1088/0004-637X/699/1/486}, \href {https://ui.adsabs.harvard.edu/abs/2009ApJ...699..486C} {699, 486}

\bibitem[\protect\citeauthoryear{{Conroy}, {White}  \& {Gunn}}{{Conroy} et~al.}{2010}]{Conroy2010}
{Conroy} C.,  {White} M.,   {Gunn} J.~E.,  2010, \mn@doi [\apj] {10.1088/0004-637X/708/1/58}, \href {https://ui.adsabs.harvard.edu/abs/2010ApJ...708...58C} {708, 58}

\bibitem[\protect\citeauthoryear{{Cutri} et~al.,}{{Cutri} et~al.}{2021}]{2014Cutri}
{Cutri} R.~M.,  et~al., 2021, {VizieR Online Data Catalog: AllWISE Data Release (Cutri+ 2013)}, VizieR On-line Data Catalog: II/328. Originally published in: IPAC/Caltech (2013)

\bibitem[\protect\citeauthoryear{{D'Cruz}, {Dorman}, {Rood}  \& {O'Connell}}{{D'Cruz} et~al.}{1996}]{Dcruz1996}
{D'Cruz} N.~L.,  {Dorman} B.,  {Rood} R.~T.,   {O'Connell} R.~W.,  1996, \mn@doi [\apj] {10.1086/177515}, \href {https://ui.adsabs.harvard.edu/abs/1996ApJ...466..359D} {466, 359}

\bibitem[\protect\citeauthoryear{{Davies} et~al.,}{{Davies} et~al.}{2010}]{2010Davies_Hevics}
{Davies} J.~I.,  et~al., 2010, \mn@doi [\aap] {10.1051/0004-6361/201014571}, \href {https://ui.adsabs.harvard.edu/abs/2010A&A...518L..48D} {518, L48}

\bibitem[\protect\citeauthoryear{{Donas} et~al.,}{{Donas} et~al.}{2007}]{Donas2007}
{Donas} J.,  et~al., 2007, \mn@doi [\apjs] {10.1086/516643}, \href {https://ui.adsabs.harvard.edu/abs/2007ApJS..173..597D} {173, 597}

\bibitem[\protect\citeauthoryear{{Dorman}, {O'Connell}  \& {Rood}}{{Dorman} et~al.}{1995}]{Dorman1995}
{Dorman} B.,  {O'Connell} R.~W.,   {Rood} R.~T.,  1995, \mn@doi [\apj] {10.1086/175428}, \href {https://ui.adsabs.harvard.edu/abs/1995ApJ...442..105D} {442, 105}

\bibitem[\protect\citeauthoryear{{Dorman}, {O'Connell}  \& {Rood}}{{Dorman} et~al.}{2003}]{Dorman2003}
{Dorman} B.,  {O'Connell} R.~W.,   {Rood} R.~T.,  2003, \mn@doi [\apj] {10.1086/375413}, \href {https://ui.adsabs.harvard.edu/abs/2003ApJ...591..878D} {591, 878}

\bibitem[\protect\citeauthoryear{{Draine} \& {Li}}{{Draine} \& {Li}}{2007}]{Draine2007}
{Draine} B.~T.,  {Li} A.,  2007, \mn@doi [\apj] {10.1086/511055}, \href {https://ui.adsabs.harvard.edu/abs/2007ApJ...657..810D} {657, 810}

\bibitem[\protect\citeauthoryear{{Draine} et~al.,}{{Draine} et~al.}{2014}]{2014Draine}
{Draine} B.~T.,  et~al., 2014, \mn@doi [\apj] {10.1088/0004-637X/780/2/172}, \href {https://ui.adsabs.harvard.edu/abs/2014ApJ...780..172D} {780, 172}

\bibitem[\protect\citeauthoryear{{Ebeling}, {Edge}, {Bohringer}, {Allen}, {Crawford}, {Fabian}, {Voges}  \& {Huchra}}{{Ebeling} et~al.}{1998}]{Ebeling1998}
{Ebeling} H.,  {Edge} A.~C.,  {Bohringer} H.,  {Allen} S.~W.,  {Crawford} C.~S.,  {Fabian} A.~C.,  {Voges} W.,   {Huchra} J.~P.,  1998, \mn@doi [\mnras] {10.1046/j.1365-8711.1998.01949.x}, \href {https://ui.adsabs.harvard.edu/abs/1998MNRAS.301..881E} {301, 881}

\bibitem[\protect\citeauthoryear{{Edler}, {Roberts}, {Boselli}, {de Gasperin}, {Heesen}, {Br{\"u}ggen}, {Ignesti}  \& {Gajovi{\'c}}}{{Edler} et~al.}{2024}]{2024Edler}
{Edler} H.~W.,  {Roberts} I.~D.,  {Boselli} A.,  {de Gasperin} F.,  {Heesen} V.,  {Br{\"u}ggen} M.,  {Ignesti} A.,   {Gajovi{\'c}} L.,  2024, \mn@doi [\aap] {10.1051/0004-6361/202348301}, \href {https://ui.adsabs.harvard.edu/abs/2024A&A...683A.149E} {683, A149}

\bibitem[\protect\citeauthoryear{{Evans}, {Parker}  \& {Roberts}}{{Evans} et~al.}{2018}]{Evans2018}
{Evans} F.~A.,  {Parker} L.~C.,   {Roberts} I.~D.,  2018, \mn@doi [\mnras] {10.1093/mnras/sty581}, \href {https://ui.adsabs.harvard.edu/abs/2018MNRAS.476.5284E} {476, 5284}

\bibitem[\protect\citeauthoryear{{Ferguson} \& {Davidsen}}{{Ferguson} \& {Davidsen}}{1992}]{Ferguson1992}
{Ferguson} H.~C.,  {Davidsen} A.~F.,  1992, in American Astronomical Society Meeting Abstracts. p. 82.04

\bibitem[\protect\citeauthoryear{Foreman-Mackey, Sick  \& Johnson}{Foreman-Mackey et~al.}{2014}]{danforemanmackey2014}
Foreman-Mackey D.,  Sick J.,   Johnson B.,  2014, python-fsps: Python bindings to FSPS (v0.1.1), \mn@doi{10.5281/zenodo.12157}, \url {https://doi.org/10.5281/zenodo.12157}

\bibitem[\protect\citeauthoryear{{Gao}, {Li}  \& {Xue}}{{Gao} et~al.}{2019}]{Gao2018}
{Gao} F.-Y.,  {Li} J.-Y.,   {Xue} Y.-Q.,  2019, \mn@doi [Research in Astronomy and Astrophysics] {10.1088/1674-4527/19/3/39}, \href {https://ui.adsabs.harvard.edu/abs/2019RAA....19...39G} {19, 039}

\bibitem[\protect\citeauthoryear{{Gil de Paz} et~al.,}{{Gil de Paz} et~al.}{2007}]{2007gildepaz}
{Gil de Paz} A.,  et~al., 2007, \mn@doi [\apjs] {10.1086/516636}, \href {https://ui.adsabs.harvard.edu/abs/2007ApJS..173..185G} {173, 185}

\bibitem[\protect\citeauthoryear{{Goudfrooij}}{{Goudfrooij}}{2018}]{2018Goudfrooij}
{Goudfrooij} P.,  2018, \mn@doi [\apj] {10.3847/1538-4357/aab553}, \href {https://ui.adsabs.harvard.edu/abs/2018ApJ...857...16G} {857, 16}

\bibitem[\protect\citeauthoryear{{Greggio} \& {Renzini}}{{Greggio} \& {Renzini}}{1990}]{Greggioandrezini1990}
{Greggio} L.,  {Renzini} A.,  1990, \mn@doi [\apj] {10.1086/169384}, \href {https://ui.adsabs.harvard.edu/abs/1990ApJ...364...35G} {364, 35}

\bibitem[\protect\citeauthoryear{{Han}, {Podsiadlowski}  \& {Lynas-Gray}}{{Han} et~al.}{2007}]{Han2007}
{Han} Z.,  {Podsiadlowski} P.,   {Lynas-Gray} A.~E.,  2007, \mn@doi [\mnras] {10.1111/j.1365-2966.2007.12151.x}, \href {https://ui.adsabs.harvard.edu/abs/2007MNRAS.380.1098H} {380, 1098}

\bibitem[\protect\citeauthoryear{{Helou}, {Madore}, {Schmitz}, {Bicay}, {Wu}  \& {Bennett}}{{Helou} et~al.}{1991}]{NED1991}
{Helou} G.,  {Madore} B.~F.,  {Schmitz} M.,  {Bicay} M.~D.,  {Wu} X.,   {Bennett} J.,  1991, {The NASA/IPAC extragalactic database.}.
pp 89--106, \mn@doi{10.1007/978-94-011-3250-3_10}

\bibitem[\protect\citeauthoryear{{Hern{\'a}ndez-P{\'e}rez} \& {Bruzual}}{{Hern{\'a}ndez-P{\'e}rez} \& {Bruzual}}{2014}]{2014perezandbruzual}
{Hern{\'a}ndez-P{\'e}rez} F.,  {Bruzual} G.,  2014, \mn@doi [\mnras] {10.1093/mnras/stu1627}, \href {https://ui.adsabs.harvard.edu/abs/2014MNRAS.444.2571H} {444, 2571}

\bibitem[\protect\citeauthoryear{Hernández-Pérez \& Bruzual}{Hernández-Pérez \& Bruzual}{2014}]{Hernandez-perez2014}
Hernández-Pérez F.,  Bruzual G.,  2014, \mn@doi [Monthly Notices of the Royal Astronomical Society] {10.1093/mnras/stu1627}, 444, 2571

\bibitem[\protect\citeauthoryear{{Hunt} et~al.,}{{Hunt} et~al.}{2019}]{Hunt2019}
{Hunt} L.~K.,  et~al., 2019, \mn@doi [\aap] {10.1051/0004-6361/201834212}, \href {https://ui.adsabs.harvard.edu/abs/2019A&A...621A..51H} {621, A51}

\bibitem[\protect\citeauthoryear{{Jarrett}, {Chester}, {Cutri}, {Schneider}  \& {Huchra}}{{Jarrett} et~al.}{2003}]{2003Jarrett}
{Jarrett} T.~H.,  {Chester} T.,  {Cutri} R.,  {Schneider} S.~E.,   {Huchra} J.~P.,  2003, \mn@doi [\aj] {10.1086/345794}, \href {https://ui.adsabs.harvard.edu/abs/2003AJ....125..525J} {125, 525}

\bibitem[\protect\citeauthoryear{{Kaviraj} et~al.,}{{Kaviraj} et~al.}{2007}]{Kaviraj2007}
{Kaviraj} S.,  et~al., 2007, \mn@doi [\apjs] {10.1086/516633}, \href {https://ui.adsabs.harvard.edu/abs/2007ApJS..173..619K} {173, 619}

\bibitem[\protect\citeauthoryear{{Kennicutt}}{{Kennicutt}}{1998}]{1998Kennicutt}
{Kennicutt} Robert~C. J.,  1998, \mn@doi [\araa] {10.1146/annurev.astro.36.1.189}, \href {https://ui.adsabs.harvard.edu/abs/1998ARA&A..36..189K} {36, 189}

\bibitem[\protect\citeauthoryear{{Kim} et~al.,}{{Kim} et~al.}{2014}]{Kim2014}
{Kim} S.,  et~al., 2014, \mn@doi [\apjs] {10.1088/0067-0049/215/2/22}, \href {https://ui.adsabs.harvard.edu/abs/2014ApJS..215...22K} {215, 22}

\bibitem[\protect\citeauthoryear{{Maraston}}{{Maraston}}{1998}]{1998Maraston}
{Maraston} C.,  1998, \mn@doi [\mnras] {10.1046/j.1365-8711.1998.01947.x}, \href {https://ui.adsabs.harvard.edu/abs/1998MNRAS.300..872M} {300, 872}

\bibitem[\protect\citeauthoryear{{Maraston}}{{Maraston}}{2005}]{Maraston2005}
{Maraston} C.,  2005, \mn@doi [\mnras] {10.1111/j.1365-2966.2005.09270.x}, \href {https://ui.adsabs.harvard.edu/abs/2005MNRAS.362..799M} {362, 799}

\bibitem[\protect\citeauthoryear{{Maraston}, {Daddi}, {Renzini}, {Cimatti}, {Dickinson}, {Papovich}, {Pasquali}  \& {Pirzkal}}{{Maraston} et~al.}{2006}]{2006AMaraston}
{Maraston} C.,  {Daddi} E.,  {Renzini} A.,  {Cimatti} A.,  {Dickinson} M.,  {Papovich} C.,  {Pasquali} A.,   {Pirzkal} N.,  2006, \mn@doi [\apj] {10.1086/508143}, \href {https://ui.adsabs.harvard.edu/abs/2006ApJ...652...85M} {652, 85}

\bibitem[\protect\citeauthoryear{{Martin} et~al.,}{{Martin} et~al.}{2005}]{Martin2005}
{Martin} D.~C.,  et~al., 2005, \mn@doi [\apjl] {10.1086/426387}, \href {https://ui.adsabs.harvard.edu/abs/2005ApJ...619L...1M} {619, L1}

\bibitem[\protect\citeauthoryear{{Mei} et~al.,}{{Mei} et~al.}{2007}]{2007Mei}
{Mei} S.,  et~al., 2007, \mn@doi [\apj] {10.1086/509598}, \href {https://ui.adsabs.harvard.edu/abs/2007ApJ...655..144M} {655, 144}

\bibitem[\protect\citeauthoryear{{Noll}, {Burgarella}, {Giovannoli}, {Buat}, {Marcillac}  \& {Mu{\~n}oz-Mateos}}{{Noll} et~al.}{2009}]{Noll2009}
{Noll} S.,  {Burgarella} D.,  {Giovannoli} E.,  {Buat} V.,  {Marcillac} D.,   {Mu{\~n}oz-Mateos} J.~C.,  2009, \mn@doi [\aap] {10.1051/0004-6361/200912497}, \href {https://ui.adsabs.harvard.edu/abs/2009A&A...507.1793N} {507, 1793}

\bibitem[\protect\citeauthoryear{{O'Connell}}{{O'Connell}}{1999a}]{OConnell1999}
{O'Connell} R.~W.,  1999a, \mn@doi [\araa] {10.1146/annurev.astro.37.1.603}, \href {https://ui.adsabs.harvard.edu/abs/1999ARA&A..37..603O} {37, 603}

\bibitem[\protect\citeauthoryear{{O'Connell}}{{O'Connell}}{1999b}]{O'Connell1999}
{O'Connell} R.~W.,  1999b, \mn@doi [\araa] {10.1146/annurev.astro.37.1.603}, \href {https://ui.adsabs.harvard.edu/abs/1999ARA&A..37..603O} {37, 603}

\bibitem[\protect\citeauthoryear{{Oconnell}, {Thuan}  \& {Puschell}}{{Oconnell} et~al.}{1986}]{Oconnel1986}
{Oconnell} R.~W.,  {Thuan} T.~X.,   {Puschell} J.~J.,  1986, \mn@doi [\apjl] {10.1086/184648}, \href {https://ui.adsabs.harvard.edu/abs/1986ApJ...303L..37O} {303, L37}

\bibitem[\protect\citeauthoryear{{Ohl} et~al.,}{{Ohl} et~al.}{1998}]{1998ohl}
{Ohl} R.~G.,  et~al., 1998, \mn@doi [\apjl] {10.1086/311605}, \href {https://ui.adsabs.harvard.edu/abs/1998ApJ...505L..11O} {505, L11}

\bibitem[\protect\citeauthoryear{{Owen}, {Eilek}  \& {Kassim}}{{Owen} et~al.}{2000}]{2000Owen}
{Owen} F.~N.,  {Eilek} J.~A.,   {Kassim} N.~E.,  2000, \mn@doi [\apj] {10.1086/317151}, \href {https://ui.adsabs.harvard.edu/abs/2000ApJ...543..611O} {543, 611}

\bibitem[\protect\citeauthoryear{{Pastorello}, {Sarzi}, {Cappellari}, {Emsellem}, {Mamon}, {Bacon}, {Davies}  \& {de Zeeuw}}{{Pastorello} et~al.}{2013}]{2013Pastorello}
{Pastorello} N.,  {Sarzi} M.,  {Cappellari} M.,  {Emsellem} E.,  {Mamon} G.~A.,  {Bacon} R.,  {Davies} R.~L.,   {de Zeeuw} P.~T.,  2013, \mn@doi [\mnras] {10.1093/mnras/sts691}, \href {https://ui.adsabs.harvard.edu/abs/2013MNRAS.430.1219P} {430, 1219}

\bibitem[\protect\citeauthoryear{{Paturel}, {Petit}, {Prugniel}, {Theureau}, {Rousseau}, {Brouty}, {Dubois}  \& {Cambr{\'e}sy}}{{Paturel} et~al.}{2003}]{Hyperleda2003}
{Paturel} G.,  {Petit} C.,  {Prugniel} P.,  {Theureau} G.,  {Rousseau} J.,  {Brouty} M.,  {Dubois} P.,   {Cambr{\'e}sy} L.,  2003, \mn@doi [\aap] {10.1051/0004-6361:20031411}, \href {https://ui.adsabs.harvard.edu/abs/2003A&A...412...45P} {412, 45}

\bibitem[\protect\citeauthoryear{{Peng} \& {Nagai}}{{Peng} \& {Nagai}}{2009}]{peng&nagai2009}
{Peng} F.,  {Nagai} D.,  2009, \mn@doi [\apjl] {10.1088/0004-637X/705/1/L58}, \href {https://ui.adsabs.harvard.edu/abs/2009ApJ...705L..58P} {705, L58}

\bibitem[\protect\citeauthoryear{{Phillipps} et~al.,}{{Phillipps} et~al.}{2020}]{philips2020}
{Phillipps} S.,  et~al., 2020, \mn@doi [\mnras] {10.1093/mnras/stz3552}, \href {https://ui.adsabs.harvard.edu/abs/2020MNRAS.492.2128P} {492, 2128}

\bibitem[\protect\citeauthoryear{{Renzini}}{{Renzini}}{2006}]{Renzini2006}
{Renzini} A.,  2006, \mn@doi [\araa] {10.1146/annurev.astro.44.051905.092450}, \href {https://ui.adsabs.harvard.edu/abs/2006ARA&A..44..141R} {44, 141}

\bibitem[\protect\citeauthoryear{{Salim} et~al.,}{{Salim} et~al.}{2016}]{2016Salim}
{Salim} S.,  et~al., 2016, \mn@doi [\apjs] {10.3847/0067-0049/227/1/2}, \href {https://ui.adsabs.harvard.edu/abs/2016ApJS..227....2S} {227, 2}

\bibitem[\protect\citeauthoryear{{Salim}, {Boquien}  \& {Lee}}{{Salim} et~al.}{2018}]{2018Salim}
{Salim} S.,  {Boquien} M.,   {Lee} J.~C.,  2018, \mn@doi [\apj] {10.3847/1538-4357/aabf3c}, \href {https://ui.adsabs.harvard.edu/abs/2018ApJ...859...11S} {859, 11}

\bibitem[\protect\citeauthoryear{{Salom{\'e}} \& {Combes}}{{Salom{\'e}} \& {Combes}}{2008}]{2008salome_combes}
{Salom{\'e}} P.,  {Combes} F.,  2008, \mn@doi [\aap] {10.1051/0004-6361:200810262}, \href {https://ui.adsabs.harvard.edu/abs/2008A&A...489..101S} {489, 101}

\bibitem[\protect\citeauthoryear{{S{\'a}nchez Almeida}, {Terlevich}, {Terlevich}, {Cid Fernandes}  \& {Morales-Luis}}{{S{\'a}nchez Almeida} et~al.}{2012}]{Almeida2012}
{S{\'a}nchez Almeida} J.,  {Terlevich} R.,  {Terlevich} E.,  {Cid Fernandes} R.,   {Morales-Luis} A.~B.,  2012, \mn@doi [\apj] {10.1088/0004-637X/756/2/163}, \href {https://ui.adsabs.harvard.edu/abs/2012ApJ...756..163S} {756, 163}

\bibitem[\protect\citeauthoryear{{Schawinski} et~al.,}{{Schawinski} et~al.}{2007}]{Schawinski2007}
{Schawinski} K.,  et~al., 2007, \mn@doi [\apjs] {10.1086/516631}, \href {https://ui.adsabs.harvard.edu/abs/2007ApJS..173..512S} {173, 512}

\bibitem[\protect\citeauthoryear{{Schombert}}{{Schombert}}{2016}]{Schombert2016}
{Schombert} J.~M.,  2016, \mn@doi [\aj] {10.3847/0004-6256/152/6/214}, \href {https://ui.adsabs.harvard.edu/abs/2016AJ....152..214S} {152, 214}

\bibitem[\protect\citeauthoryear{{Sheen}, {Yi}, {Ree}, {Jaff{\'e}}, {Demarco}  \& {Treister}}{{Sheen} et~al.}{2016}]{sheen2016}
{Sheen} Y.-K.,  {Yi} S.~K.,  {Ree} C.~H.,  {Jaff{\'e}} Y.,  {Demarco} R.,   {Treister} E.,  2016, \mn@doi [\apj] {10.3847/0004-637X/827/1/32}, \href {https://ui.adsabs.harvard.edu/abs/2016ApJ...827...32S} {827, 32}

\bibitem[\protect\citeauthoryear{{Skrutskie} et~al.,}{{Skrutskie} et~al.}{2003}]{2003Skrutskie}
{Skrutskie} M.~F.,  et~al., 2003, {VizieR Online Data Catalog: The 2MASS Extended sources (IPAC/UMass, 2003-2006)}, VizieR On-line Data Catalog: VII/233. Originally published in: 2006AJ....131.1163S

\bibitem[\protect\citeauthoryear{{Skrutskie} et~al.,}{{Skrutskie} et~al.}{2006}]{20062MASS}
{Skrutskie} M.~F.,  et~al., 2006, \mn@doi [\aj] {10.1086/498708}, \href {https://ui.adsabs.harvard.edu/abs/2006AJ....131.1163S} {131, 1163}

\bibitem[\protect\citeauthoryear{{Smith}, {Lucey}  \& {Carter}}{{Smith} et~al.}{2012}]{Smith2012}
{Smith} R.~J.,  {Lucey} J.~R.,   {Carter} D.,  2012, \mn@doi [\mnras] {10.1111/j.1365-2966.2012.20524.x}, \href {https://ui.adsabs.harvard.edu/abs/2012MNRAS.421.2982S} {421, 2982}

\bibitem[\protect\citeauthoryear{{Sohn}, {O'Connell}, {Kundu}, {Landsman}, {Burstein}, {Bohlin}, {Frogel}  \& {Rose}}{{Sohn} et~al.}{2006}]{2006Sohn}
{Sohn} S.~T.,  {O'Connell} R.~W.,  {Kundu} A.,  {Landsman} W.~B.,  {Burstein} D.,  {Bohlin} R.~C.,  {Frogel} J.~A.,   {Rose} J.~A.,  2006, \mn@doi [\aj] {10.1086/499039}, \href {https://ui.adsabs.harvard.edu/abs/2006AJ....131..866S} {131, 866}

\bibitem[\protect\citeauthoryear{{Thomas}, {Maraston}, {Bender}  \& {Mendes de Oliveira}}{{Thomas} et~al.}{2005}]{Thomas2005}
{Thomas} D.,  {Maraston} C.,  {Bender} R.,   {Mendes de Oliveira} C.,  2005, \mn@doi [\apj] {10.1086/426932}, \href {https://ui.adsabs.harvard.edu/abs/2005ApJ...621..673T} {621, 673}

\bibitem[\protect\citeauthoryear{{V{\'e}ron-Cetty} \& {V{\'e}ron}}{{V{\'e}ron-Cetty} \& {V{\'e}ron}}{2006}]{2006VCC1316AGN}
{V{\'e}ron-Cetty} M.~P.,  {V{\'e}ron} P.,  2006, \mn@doi [\aap] {10.1051/0004-6361:20065177}, \href {https://ui.adsabs.harvard.edu/abs/2006A&A...455..773V} {455, 773}

\bibitem[\protect\citeauthoryear{{Wright} et~al.,}{{Wright} et~al.}{2010}]{wrightwise2010}
{Wright} E.~L.,  et~al., 2010, \mn@doi [\aj] {10.1088/0004-6256/140/6/1868}, \href {https://ui.adsabs.harvard.edu/abs/2010AJ....140.1868W} {140, 1868}

\bibitem[\protect\citeauthoryear{{Yi}, {Demarque}  \& {Kim}}{{Yi} et~al.}{1997}]{Yi1997}
{Yi} S.,  {Demarque} P.,   {Kim} Y.-C.,  1997, \mn@doi [\apj] {10.1086/304192}, \href {https://ui.adsabs.harvard.edu/abs/1997ApJ...482..677Y} {482, 677}

\bibitem[\protect\citeauthoryear{{Yi} et~al.,}{{Yi} et~al.}{2005}]{Yi2005}
{Yi} S.~K.,  et~al., 2005, \mn@doi [\apjl] {10.1086/422811}, \href {https://ui.adsabs.harvard.edu/abs/2005ApJ...619L.111Y} {619, L111}

\bibitem[\protect\citeauthoryear{{Yi}, {Lee}, {Sheen}, {Jeong}, {Suh}  \& {Oh}}{{Yi} et~al.}{2011}]{Yi2011}
{Yi} S.~K.,  {Lee} J.,  {Sheen} Y.-K.,  {Jeong} H.,  {Suh} H.,   {Oh} K.,  2011, \mn@doi [\apjs] {10.1088/0067-0049/195/2/22}, \href {https://ui.adsabs.harvard.edu/abs/2011ApJS..195...22Y} {195, 22}

\bibitem[\protect\citeauthoryear{{Yoon}, {Lee}, {Rey}, {Ree}  \& {Yi}}{{Yoon} et~al.}{2004}]{Yoon2004}
{Yoon} S.-J.,  {Lee} Y.-W.,  {Rey} S.-C.,  {Ree} C.~H.,   {Yi} S.~K.,  2004, \mn@doi [\apss] {10.1023/B:ASTR.0000044325.31072.4d}, \href {https://ui.adsabs.harvard.edu/abs/2004Ap&SS.291..223Y} {291, 223}

\bibitem[\protect\citeauthoryear{{York} et~al.,}{{York} et~al.}{2000}]{SDSS2000}
{York} D.~G.,  et~al., 2000, \mn@doi [\aj] {10.1086/301513}, \href {https://ui.adsabs.harvard.edu/abs/2000AJ....120.1579Y} {120, 1579}

\bibitem[\protect\citeauthoryear{{da Cunha}, {Charlot}  \& {Elbaz}}{{da Cunha} et~al.}{2008}]{daCunha2008}
{da Cunha} E.,  {Charlot} S.,   {Elbaz} D.,  2008, \mn@doi [\mnras] {10.1111/j.1365-2966.2008.13535.x}, \href {https://ui.adsabs.harvard.edu/abs/2008MNRAS.388.1595D} {388, 1595}

\makeatother
\end{thebibliography}

%%%%%%%%%%%%%%%%%%%%%%%%%%%%%%%%%%%%%%%%%%%%%%%%%%

%%%%%%%%%%%%%%%%% APPENDICES %%%%%%%%%%%%%%%%%%%%%

\appendix
\section{Additional Information}
\label{appendix:A1}

\begin{figure}
    \centering
    \includegraphics[width=1.3\columnwidth]{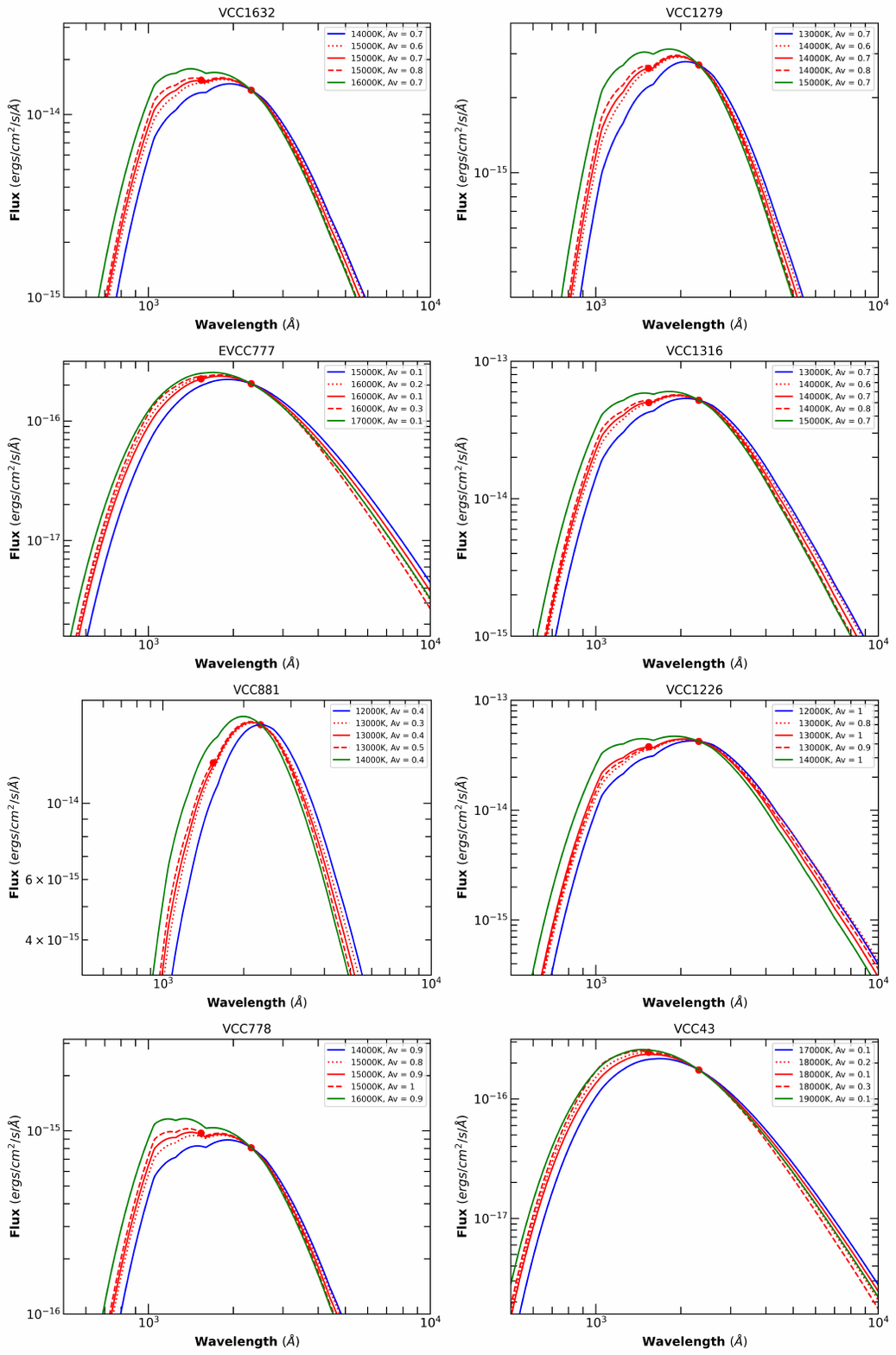}
    \caption{shows the blackbody fit for the UV flux in all the UV upturn galaxies. Each plot features blue, green, and red lines representing different temperatures. The red lines and the red dotted lines represent the best-fitted temperature with different Av values. Red line represents the best-fitted blackbody of each of the sample galaxies.}
    \label{fig:BBfit_all}
\end{figure}

\end{document}